\documentclass[twocolumn,showpacs,preprintnumbers, amsmath,amssymb,amsfonts, final, a4paper,aps,
citeautoscript,footinbib,prl,reprint,superscriptaddress]{revtex4-1}
\usepackage{times}
\usepackage{graphicx}
\usepackage[latin1]{inputenc}
\usepackage{mathrsfs}
\usepackage[dvipsnames,usenames]{color}
\usepackage{soul}
\usepackage[colorlinks,bookmarks=true,citecolor=blue,linkcolor=blue,urlcolor=blue, breaklinks=true]{hyperref}
\usepackage[vcentermath]{youngtab}
\graphicspath{{./Figs/}}

\usepackage{amssymb}
\usepackage{latexsym}
\usepackage{amsmath}
\usepackage{epsfig}
\usepackage{xr}
\definecolor{darkred}{rgb}{0.8,0.1,0.1}
\definecolor{purple}{rgb}{0.45,0.1,0.45}
\hypersetup{colorlinks=true, linkcolor=darkred, citecolor=blue}
\newcommand{\VEV}[1]{\langle #1 \rangle}  
\newcommand{\be}{\begin{equation}}
\newcommand{\ee}{\end{equation}}
\newcommand{\bea}{\begin{eqnarray}}
\newcommand{\eea}{\end{eqnarray}}

\renewcommand{\vec}[1]{\boldsymbol{#1}} 
\begin{document}
\title{%
``Haldane'' phases with ultracold fermionic atoms in double-well optical lattices}
\author{P. Fromholz} 
\affiliation{Laboratoire de Physique Th\'eorique et Mod\'elisation,
  CNRS UMR 8089, Universit\'e de Cergy-Pontoise, Site de Saint-Martin,
  F-95300 Cergy-Pontoise Cedex, France}
\author{S. Capponi} 
\affiliation{Laboratoire de Physique Th\'eorique, CNRS UMR 5152,
  Universit\'e Paul Sabatier, F-31062 Toulouse, France.}
\author{P. Lecheminant} 
\affiliation{Laboratoire de Physique Th\'eorique et Mod\'elisation,
  CNRS UMR 8089, Universit\'e de Cergy-Pontoise, Site de Saint-Martin,
  F-95300 Cergy-Pontoise Cedex, France}
\author{D. J. Papoular} 
\affiliation{Laboratoire de Physique Th\'eorique et Mod\'elisation,
  CNRS UMR 8089, Universit\'e de Cergy-Pontoise, Site de Saint-Martin,
  F-95300 Cergy-Pontoise Cedex, France}
\author{K. Totsuka} 
\affiliation{Yukawa Institute for Theoretical Physics, 
Kyoto University, Kitashirakawa Oiwake-Cho, Kyoto 606-8502, Japan.}
\date{\today}
\begin{abstract}
We propose to realize  one-dimensional topological phases protected by 
SU($N$) symmetry  using alkali or alkaline-earth atoms  loaded into a
bichromatic optical  lattice.  We  derive a  realistic model  for this
system and investigate  it theoretically.  
Depending on the parity of $N$, 
two different classes of symmetry-protected topological (SPT) phases are stabilized at half-filling 
for physical parameters of the model. 
For even $N$,  the celebrated spin-1 Haldane phase and its generalization to SU($N$) are obtained 
with no local symmetry breaking. In  stark  contrast,  at  least  for $N=3$, a new class of SPT phases,  
dubbed chiral  Haldane phases, that spontaneously break inversion symmetry,  
emerge with a two-fold ground-state degeneracy.   
The latter ground  states  with  open-boundary conditions are characterized by
different left and right  boundary spins which are related  by
conjugation. Our results show that topological phases are within close 
reach of the latest experiments on cold fermions in optical lattices.
\end{abstract}

\maketitle

\emph{Introduction --}
Symmetry protected topological (SPT) phases have attracted 
a lot of attention over recent years.
These new quantum phases exhibit short-range entanglement and possess only conventional gapped excitations in the bulk, 
while hosting non-trivial symmetry-protected surface states 
\cite{wenbook,senthilreview}.
A paradigmatic example of one-dimensional (1D) bosonic SPT phases is the Haldane phase 
found in the spin-1 antiferromagnetic spin chain \cite{haldane}.
In the bulk, the phase looks ordinary, but, in the case of an open-boundary condition \cite{Kennedy-90} 
or when the chain is cut by doping impurities \cite{hagiwara}, non-trivial spin-1/2 edge states appear.  
This phase is protected by the SO(3) symmetry
underlying the Heisenberg model, and more generally, 
by at least one of the three discrete symmetries: the dihedral group of $\pi$-rotations along the $x ,y ,z$ axes, time-reversal or inversion symmetries \cite{gu,pollmann}.

A fairly complete understanding of 1D bosonic SPT phases has been obtained 
through various approaches such as group cohomology, matrix-product states, entanglement spectroscopy, 
and field-theoretical arguments \cite{chenx,schuch,fidowski,Pollmann2010,xu}.
The possible 1D SPT phases associated with
a given protecting symmetry $G$ are classified by its projective representations, i.e., 
the second cohomology group ${\cal H}^{2}(G, \text{U(1)})$.   
For instance, in the presence of SO(3) symmetry,  there is a $\mathbb{Z}_2$ classification and the Haldane phase is 
the only SPT phase whose edge states obey a non-trivial projective representation \cite{gu,pollmann}.

Richer SPT phases can be obtained when $G$ is a more general Lie group.  
For instance, the group SU($N$) leads to a $\mathbb{Z}_N$ classification 
predicting $N-1$ non-trivial SPT phases \cite{Duivenvoorden-Q-13} protected 
by SU($N$) (PSU($N$), more precisely \cite{comment}) or by its discrete subgroup $\mathbb{Z}_N$  $\times$ $\mathbb{Z}_N$ \cite{Else-B-D-13,Duivenvoorden-Q-ZnxZn-13}.
The edge states of these SPT phases are labeled by the inequivalent projective representations 
of SU($N$) which are specified by $\mathbb{Z}_N$ quantum numbers $n_{\rm top} = n_{\text{Y}} ({\rm mod}\;  N)$,
with $n_{\text{Y}}$ being the number of boxes in the Young diagram corresponding to the representation
of the boundary spins \cite{Duivenvoorden-Q-13,Capponi-L-T-15}. 
In stark contrast to the $N=2$ case, i.e. $G= \text{SO(3)}$,  
where all the projective representations are self-conjugate, the left and right edge states of the SU($N$) SPT phases 
with $N>2$ might belong to {\em different} projective representations that are related by conjugation.  
This leads to an interesting class of SPT phases, dubbed chiral Haldane ($\chi$H),  
which spontaneously break the inversion symmetry.  These phases are partially characterized by local order parameters and 
exist in pairs; 
in one phase, the left and right edge states transform respectively 
in the SU($N$) representation $\mathcal{R}$ and its conjugate $\bar{\mathcal{R}}$, and vice-versa in the other \cite{Furusaki2014,greiter,AKLT,Quella2015}.  
In the following, we label the SPT phases by the number of boxes in the Young diagrams  
as $(n_{\text{Y}}(\mathcal{R}),n_{\text{Y}}(\bar{\mathcal{R}}))$ (mod $N$).  
In reflection-symmetric systems, the two topological ground states 
$(n_{\text{Y}}(\mathcal{R}),n_{\text{Y}}(\bar{\mathcal{R}}))$ and 
$(n_{\text{Y}}(\bar{\mathcal{R}}),n_{\text{Y}}(\mathcal{R}))$ are degenerate.
\begin{figure}[htb]
  \begin{minipage}{.45\textwidth}
   \includegraphics[width=1.0\linewidth]{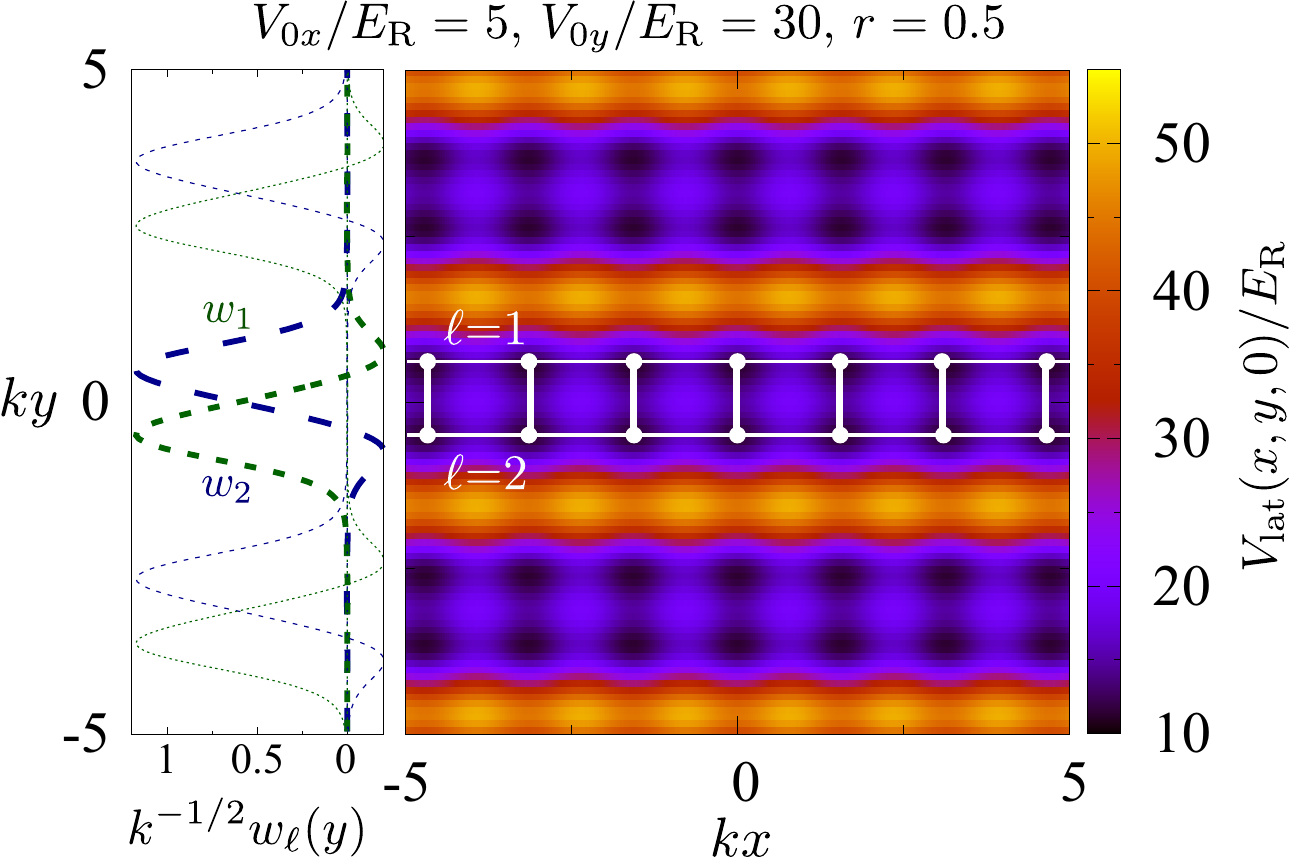}
  \end{minipage}
  \caption{ \label{fig:Vlat_wannier} (Color online) 
    The potential $V_\mathrm{lat}$ in the $xy$ plane (right panel): three double-well ladder systems are visible. 
    The two independent Wannier functions $w_{1}(y)$ and $w_{2}(y)$ along $y$ (left, green and blue) are centered 
    on the two chains ($\ell=1,2$). 
  Lengths and energies are respectively expressed in units of the 
  reduced wavelength $1/k$ and the recoil energy $E_R = \hbar^2 k^2/2m$.  
  This optical potential yields $t_{\perp}/t=2.9$ and $V/U=0.086$.}
\end{figure} 

In this Letter, we propose an implementation of the Haldane phase ($N=2$) and its generalizations 
to even-$N$, as well as the $\chi$H phases for $N=3$,
with half-filled ultracold fermions loaded into 1D double-well optical lattices. 
Thanks to their cleanness and controllability,
these systems offer an ideal framework for 
the realization of the SPT phases, which requires precise symmetries.
The $N=2$ case may be realized using the two lowest
hyperfine states of ${}^6\mathrm{Li}$. Larger values of $N$ may be explored experimentally 
using ${}^{87}\mathrm{Sr}$ or ${}^{173}\mathrm{Yb}$ atoms in their ${}^{1}S_0$ ground states, 
which possess SU($N$)-symmetry ($N \leq 10$) \cite{Cazalilla-H-U-09,Gorshkov-et-al-10,Cazalilla-R-14,Taie2012,
Pagano2014,Zhang2014,Scazza2014}.  
By means of complementary strong-coupling and numerical techniques, 
we show that, for all even $N \ge 2$ and (at least) $N=3$, fully gapped featureless Mott-insulating phases  
show up in the phase diagram of the underlying lattice fermion models with repulsive interactions.  
The phases occurring for even-$N$ are identified as the Haldane phase ($N=2$) or its generalization ($N \geq 4$).  
On the other hand, for odd $N$ (at least for $N=3$), we find that 
$\chi$H phases emerge breaking the inversion symmetry spontaneously.  
As we will see, these SPT phases are stabilized for realistic parameters of the model, 
a result which paves the way to their experimental investigation for $N \le 10$.  

\emph{Model --}
We consider ultracold (alkali, alkaline--earth, or ytterbium) fermions
with SU($N$) symmetry,
trapped inside the following potential representing a
three--dimensional array of double wells
(see Fig.~\ref{fig:Vlat_wannier}):
\begin{equation} \label{eq:latticepot}
  \begin{split}
    V_\text{lat}(x,y,z) =
    & V_{0y}
    \left[
      \sin^2(ky)+r\cos^2(2ky)
    \right] \\
    & + V_{0x}\sin^2(2kx) +V_{0z}\sin^2(2kz)
    \ ,
  \end{split}
\end{equation}
where $1/k$ denotes the reduced wavelength  and $r$ is a tunable parameter.  
This potential can be realized optically,
using a bichromatic lattice \cite{atala} or exploiting interference patterns involving two light beams
with different polarizations \cite{sebbystrabley}. 
Choosing sufficiently large values of $V_{0y}$ and $V_{0z}$,
we obtain a single 1D two-leg ladder whose legs ($\ell=1$ or $2$) and rungs (labeled $i$) 
are respectively parallel to the $x$ and $y$ axes. 

We restrict our analysis to the lowest bands  
in the $x$ and $y$ directions.  In the $y$-direction,  
we keep the two lowest bands so as to resolve the two minima 
of each double well. This leads to the following lattice model:
\begin{equation}
\begin{split}
{\cal H}_0 =& -t \sum_{i, \ell} \sum_{\alpha=1}^{N} \left( 
c_{\ell \alpha,i+1}^\dagger c_{\ell \alpha,i}+ \text{H.c.} \right) - \mu \sum_{i} n_i   \\
&- t_\perp \sum_{i} \sum_{\alpha=1}^{N} \left( c_{1\alpha,i}^\dagger c_{2 \alpha,i}+ \text{H.c.} \right),
\end{split}
  \label{eq:ham0} 
\end{equation}
where the operator $c^{\dagger}_{\ell\alpha,\,i}$ creates
a fermion in the nuclear-spin state $\alpha(=1,\cdots,N)$ on the leg $\ell(=1,2)$ and the rung $i$. 
In Eq.~(\ref{eq:ham0}), the total density operator on the rung $i$
is $n_i = \sum_{\ell\alpha}c_{\ell \alpha,i}^\dagger c_{\ell \alpha,i}=\sum_{\ell \alpha} n_{\ell \alpha, i}$,  
and the tunneling amplitudes $t$ along a leg and $t_\perp$ along a rung
are different in general.
We now account for $\operatorname{SU}(N)$-symmetric 2-body interactions modeled by the contact Hamiltonian
$g\sum_{\alpha\neq\beta}
\int d^3r \, n_\alpha(\vec{r}) n_\beta(\vec{r})$,
where $n_\alpha(\vec{r})$ is the density operator for fermions in the 
internal nuclear state $\alpha$ \cite{Cazalilla-H-U-09,Gorshkov-et-al-10,Cazalilla-R-14}.
Retaining the same bands as in Eq.~(\ref{eq:ham0}),
we obtain the following interaction Hamiltonian:
\begin{equation}
\begin{split}
{\cal H}_{\mathrm{int}} = & \frac{U}{2}\sum_{i}\sum_{\ell=1}^{2}\sum_{\alpha \neq \beta}n_{\ell  \alpha, i}n_{\ell  \beta, i}  \\
&+ V \sum_{i}\sum_{\alpha \neq \beta} \biggl\{ n_{1 \alpha,i}n_{2 \beta,i}
 +c_{1 \alpha,i}^\dagger c_{2 \beta,i}^\dagger c_{1 \beta,i}c_{2 \alpha,i} 
  \\
&+ 
 \frac{1}{2}\left(c_{1 \alpha,i}^\dagger c_{1 \beta,i}^\dagger c_{2 \beta,i}c_{2 \alpha,i} + \text{H.c.} \right) \biggr\}   \; , 
\end{split}
 \label{inthamferm}
\end{equation}
where $U$ is the on-site interaction, and $V$ encodes the off-site interaction 
between the two sites on a given rung.
There are three types of off-site processes: 
(i) density-density interaction,
(ii) spin-exchange interaction,
and (iii) pair-hopping of fermions with different spins from one leg to the other.  
Hence, Eq.~(\ref{inthamferm}) can be viewed as a generalized two-leg fermionic SU($N$) ladder model 
with pair-hopping processes.  
The coefficients $t$, $t_\perp$, $U$ and $V$ characterizing the lattice model
$\mathcal{H} =\mathcal{H}_0+\mathcal{H}_\mathrm{int}$ 
are determined by the Wannier functions corresponding to  
$V_{\text{lat}}(\mathbf{r})$ \cite{bloch:RMP2008}, which we calculate
numerically following Ref.~\cite{bloch:AAMO2006}.
In particular, along the rung direction $y$, we choose
the Wannier functions $w_1(y)$ and $w_2(y)$ to be real and 
localized on the legs $\ell=1$ and $2$, respectively (see Fig.~\ref{fig:Vlat_wannier}).  
The orthogonality of the Wannier functions requires that $w_1(y)$ and $w_2(y)$ 
have finite extent around their center with changing signs.  
The coefficient $V$ is proportional to 
$g\int d y \: w_1^2 w_2^2$, and is finite because of a non-zero overlap
between the positive functions $w_1^2$ and $w_2^2$.  
Besides the above three interactions, density-assisted hopping terms \cite{werner:PRL2005},
proportional to the integral $g\int dy\:  w_1 w_2^3$, are also present.   
However, now the sign change of the Wannier functions strongly suppresses the integral, 
so that we can safely drop them in Eq.~\eqref{inthamferm}. 
The ratios $t_\perp/t$ and $V/U$ are fixed by the optical potential $V_\text{lat}(x,y,z)$:
$t_\perp/t$ can be tuned
from $1$ to a few units by varying the parameter $r$ in Eq.~\eqref{eq:latticepot}~\cite{note1}, whereas $V/U$ is of the order of $10^{-1}$.
The ratio $U/t$ can be tuned using a magnetic Feshbach resonance
in the case of alkali atoms \cite{chin:RMP2010}, or an optical Feshbach resonance
for alkaline--earth atoms \cite{enomoto:PRL2008,taie:PRL2016}.

\emph{Strong-coupling analysis --} 
We now consider the atomic limit of the model (\ref{inthamferm}) to investigate 
the possible existence of SPT phases
in the large-$U$ limit.   
If we introduce the antisymmetric and symmetric combinations
$d_{1\alpha,i}= (c_{1 \alpha,i}-c_{2 \alpha,i})/\sqrt{2}$ and 
$d_{2 \alpha,i}= (c_{1 \alpha,i}+c_{2 \alpha,i})/\sqrt{2}$, $\mathcal{H}$ takes the form of the $p$-band model of 
Refs.~\cite{Kobayashi-O-O-Y-M-12,Kobayashi-O-O-Y-M-14,Bois-C-L-M-T-15} 
in an (effective) orbital magnetic field proportional to $t_{\perp}$:
\begin{equation}
\begin{split}
\mathcal{H}  = & -t \sum_{i,\alpha} \sum_{m=1,2} \left(d_{m\alpha,\,i}^\dag d_{m\alpha,\,i+1}+  \text{H.c.} \right)  \\
& - \left( \mu + \frac{U+V}{2} \right)  \sum_{i}  n_i    + 2 t_\perp \sum_{i} T_i^z  \\
&+  \frac{U + V}{4} \sum_i  n_i^2 + 2 V  \sum_i (T_i^z)^2  +  (U-V)  \sum_i (T_i^x)^2  
 \; ,
\end{split}
\label{pbandmodelbis}
\end{equation}
where $T_i^a = \frac{1}{2}\sum_{m,n,\alpha} d_{m \alpha,\,i}^\dag \sigma^a_{m n} d_{n \alpha,\,i} $ is the pseudo-spin operator for the orbital degrees of freedom and 
$\sigma^a (a =x,y,z)$ the Pauli matrices.  
In what follows, we restrict ourselves to half-filling (i.e., $N$ fermions per rung). 
The atomic-limit ($U,V, t_\perp \gg t$) energy spectrum of the model 
\eqref{pbandmodelbis} is characterized by the SU($N$) and the pseudo-spin ($\mathbf{T}$) irreducible representations \cite{supp}.  
For even $N$, in most part of the region $U>V>0$, the orbital pseudo-spin $\mathbf{T}$ is quenched to a singlet,  
while the SU($N$) spin is maximized into a self-conjugate representation of SU($N$) described by a Young diagram 
with two columns with lengths $N/2$ \cite{{Bois-C-L-M-T-15}}.  
To second order in $t$, the effective Hamiltonian is given by the SU($N$) Heisenberg model~\cite{Bois-C-L-M-T-15}:
\begin{equation}
{\cal H}_{\rm eff}^{\text{(even)}} = J \sum_i \sum_{A=1}^{N^2-1} {\cal S}^{A}_{i+1} {\cal S}^{A}_{i},
  \label{eq:hameffeven} 
\end{equation}
where $J= 2t^2/(U+V)$ is the spin-exchange constant, 
and ${\cal S}^{A}_{i}$ are the local SU($N$) spin operators belonging to the self-conjugate representation mentioned above.
For $N=2$, Eq.~\eqref{eq:hameffeven} reduces to 
the spin-1 Heisenberg chain, whose ground state is in the Haldane phase. 
For generic even $N$, the ground-state properties of the model  \eqref{eq:hameffeven} have recently been
investigated in detail in Refs.~\cite{Nonne-M-C-L-T-13,Bois-C-L-M-T-15,Totsuka-15,Capponi-L-T-15,Mila-16}, 
where the ground state has been identified with an SU($N$) SPT phase 
with $\mathbb{Z}_N$ quantum numbers $n_{\text{top}} = N/2$ (mod $N$) 
characterized by edge states transforming in the antisymmetric $(N/2)$-tensor representation of SU($N$). 
Remarkably, for odd $N$, the orbital degrees of freedom play a crucial role.  
To see this, let us consider the $N=3$ case and start from $U=V$ and $t_{\perp}=0$, where 
each site of a rung is occupied either by $\mathbf{3}$ (${\tiny \yng(1)}$) or $\bar{\mathbf{3}}$ (${\tiny \yng(1,1)}$) in the atomic-limit ground state.  Regarding $\mathbf{3}$ and $\bar{\mathbf{3}}$ as the two orbital states (e.g., up and down) and carrying out the second-order perturbation in $U-V$ and $t_{\perp}$, we obtain a spin-orbital effective Hamiltonian, 
which, when $U>V$, reduces to an SU(3) two-leg ladder with different spins ($\mathbf{3}$ and $\bar{\mathbf{3}}$) on 
the two legs \cite{3-3bar-ladder}.  The point is that the couplings now depend on the orbital part and, after tracing it out, the system further reduces to the two-leg ladder with diagonal interactions. We numerically investigated the model to find that the $\chi \text{H}$ phase is stabilized {\em only} when finite diagonal interactions exist \cite{3-3bar-ladder}.     
A relatively large $t_{\perp}(>0)$ freezes the orbital pseudo-spins and the diagonal couplings, that are crucial 
to the SPT phase, disappear.  In fact, both the strong-coupling expansion assuming large $t_{\perp}$ and direct numerical simulations for large enough $t_{\perp}$ found only a featureless trivial phase~\cite{supp} in agreement with the above scenario.  

\begin{figure}[!ht]
\centering
\includegraphics[width=\columnwidth,clip]{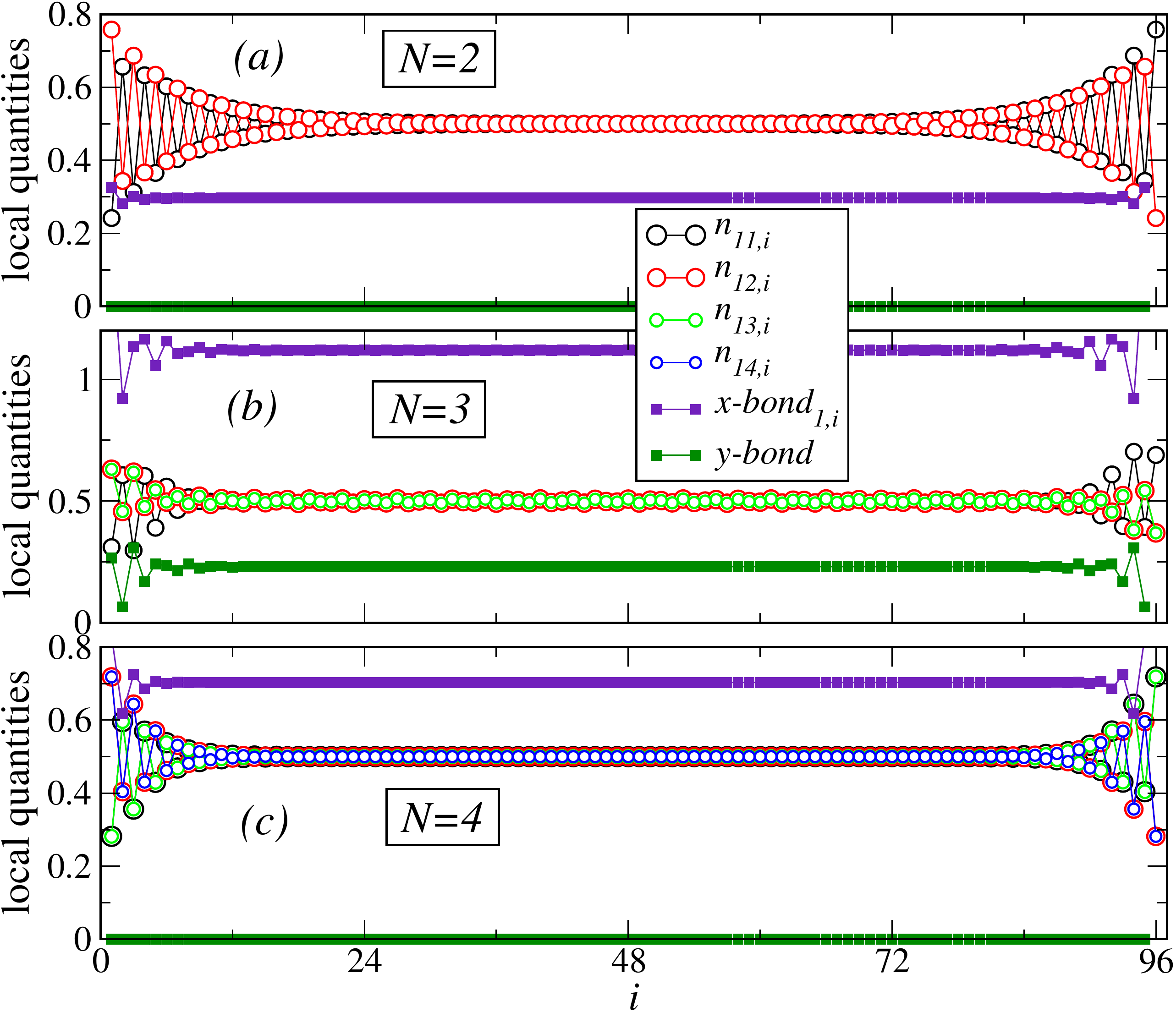} 
\caption{(Color online) Local densities and bond energies obtained by DMRG for a chain of length $L=96$ 
in the cases $N=2$, $3$ and $4$, using $t_\perp=1$, $U=12$, $V=4$.   
The densities and $x$-bond (i.e., rung) energies are found to be equal on both chains, 
and we show them for $\ell=1$. The clear evidence of edge states signals the three SPT phases.  }
\label{fig:edge}
\end{figure}

\begin{figure}[!ht]
\centering
\includegraphics[width=\columnwidth,clip]{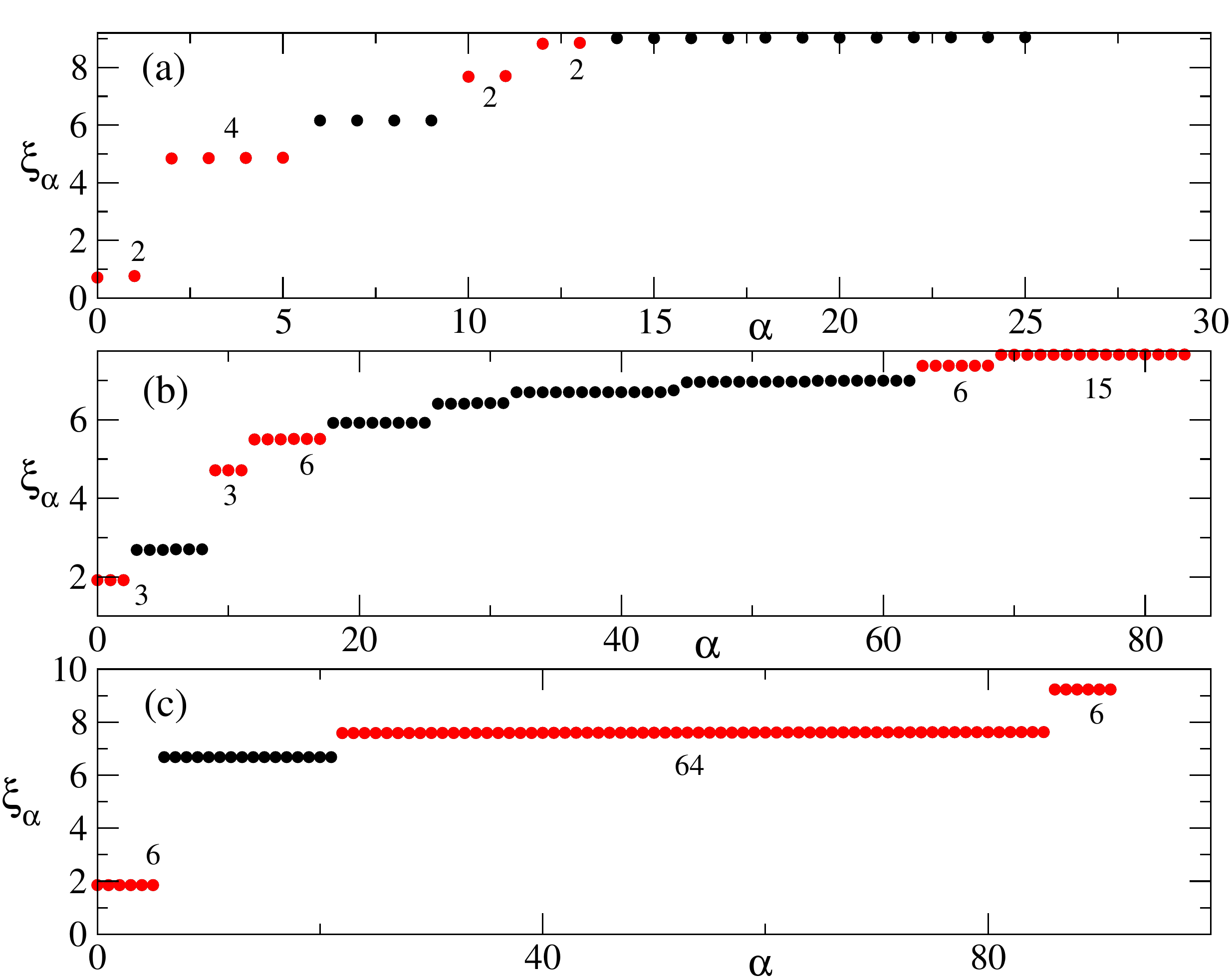}
\caption{(Color online) The ES obtained by DMRG on $L=48$ chain in the $N=2$, $3$ and $4$ model  
(from top to bottom) using $t_\perp=1$, $U=20$, $V=2$.
In all three cases, the system is in the SPT phase. Bosonic (fermionic) levels are shown by red (black) circles.  
Numbers denote the number of quasi-degenerate levels.}
\label{fig:es}
\end{figure}

\begin{figure}[!ht]
\centering
\includegraphics[width=\columnwidth,clip]{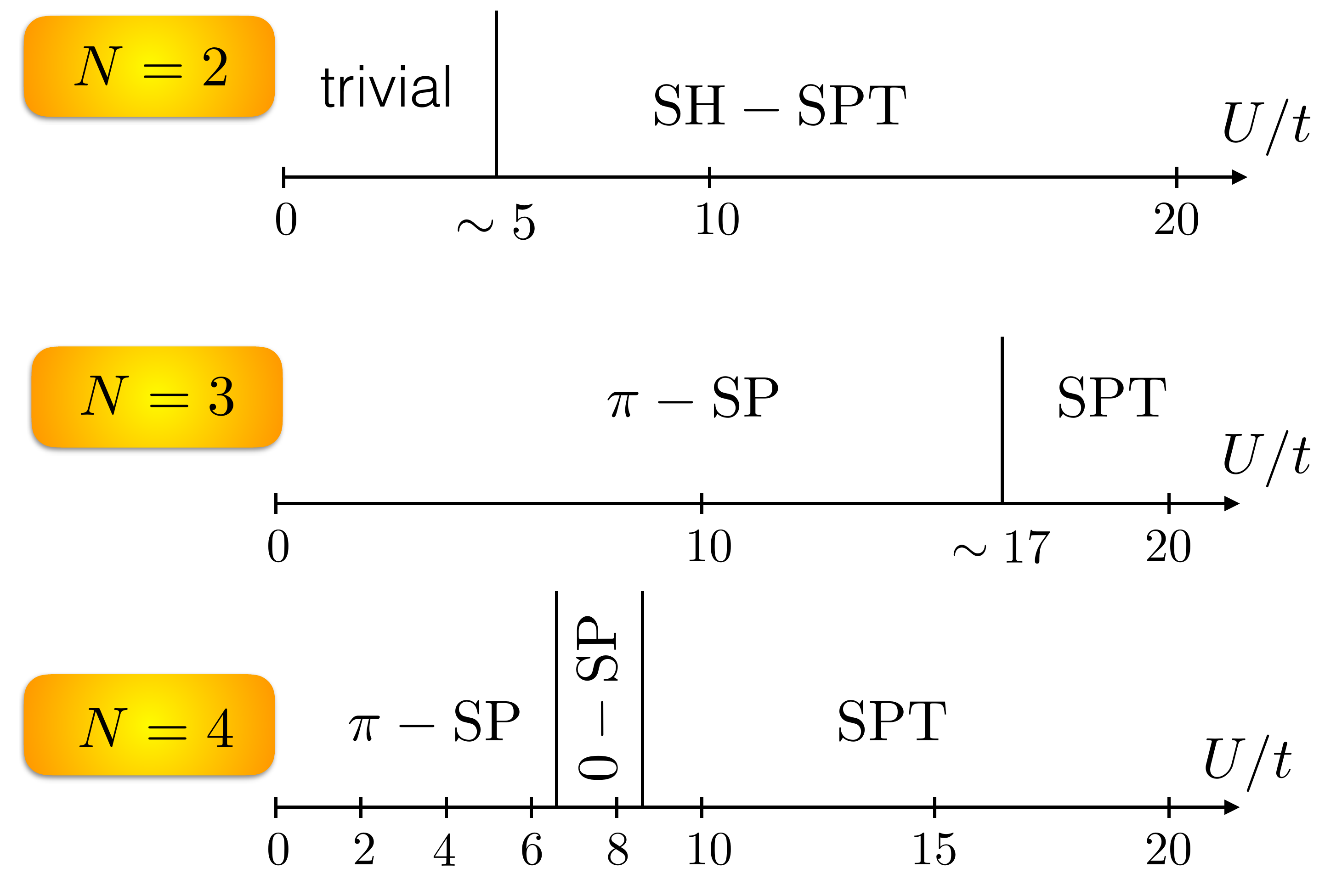} 
\caption{(Color online) Phase diagram for $N=2$, $3$ and $4$ at fixed $t_\perp=t=1$, as a function of $U$ 
(with $U/V=10$, see text) obtained from DMRG simulations.  
In all three cases, we find different SPT phases at strong coupling. 
For weaker interactions, we
find trivial non-degenerate gapped phases, 
or the out-of-phase (respectively in-phase) dimerized spin-Peierls-like $\pi$-SP (respectively $0$-SP) phase. }
\label{fig:phasediag_DMRG}
\end{figure}

\emph{Numerical calculations --}
We mapped out the zero-temperature phase diagram of the model (\ref{pbandmodelbis}) at half-filling 
by means of density-matrix renormalization-group (DMRG) calculations~\cite{DMRG}. 
We have used open boundary conditions, keeping between 2000 and 4000 states depending on the
model parameters and sizes in order to keep a discarded weight below $10^{-5}$.   
We fix $t=1$ as the unit of energy and, instead of
the full SU($N$) symmetry, we have implemented the $\text{U(1)}^{N}$ symmetry  
corresponding to the conservation of each species of fermions ($\alpha=1,\ldots,N$).  
Starting with the simplest $N=2$ case, we reveal that 
the SU($N$) SPT phases, predicted in the strong-coupling regime, persist down to realistic regions. 
Figure~\ref{fig:edge}(a) shows the presence of exponentially localized edge states 
in the spin-resolved local densities $n_{\ell \alpha, i}$, 
which is a clear signature of the spin-Haldane (SH) phase with spin--1/2 edge states. 
The possible SPT phases in the $N=3$ and $N=4$ cases can also be probed using their particular edge states [Fig.~\ref{fig:edge}(b,c)], 
or their corresponding entanglement spectra (ES) [Fig.~\ref{fig:es}(b,c)].   
The precise nature of the edge states  can be inferred from Fig.~\ref{fig:edge} 
and, for SU(3), we find that the phase for $t= t_{\perp}=1$ is a $\chi$H phase $(n_{\text{Y}}(\mathcal{R}),n_{\text{Y}}(\bar{\mathcal{R}}))=(1,2)$ with the left and right edge states respectively transforming
in the ${\bf 3}$ and ${\bf \bar 3}$ representations of SU(3) \cite{supp}.   
As has been mentioned above, when the system is inversion-symmetric, this and the second $\chi$H phase $(2,1)$ 
must be degenerate; DMRG simulations randomly pick one of the two minimally entangled states.  In fact, we found that another run with a different sweeping procedure gave access to the second one \cite{supp}.   
This signals the emergence of the $\chi$H phase $(1,2)$ or $(2,1)$ for $t= t_{\perp}=1$  
which spontaneously breaks the inversion symmetry~\cite{Furusaki2014,greiter}.  
Similarly, for $N=4$, the edge states in Fig.~\ref{fig:edge}(c) 
strongly suggest one of the three SPT phases $(2,2)$ protected by SU(4).    
Specifically, the edge states are found to belong to the self-conjugate antisymmetric representation of SU(4) 
with dimension 6, in agreement with previous studies 
on the $N=4$ case~\cite{Capponi-L-T-15,Nonne-M-C-L-T-13,Bois-C-L-M-T-15,Totsuka-15}.  
In order to provide additional insight into these SPT phases, we plot their ES obtained 
by cutting the chain in the middle and computing the Schmidt eigenvalues of the ground-state wavefunction.  
The ES of the SH phase  is known to exhibit \emph{double-degeneracy} for all levels~\cite{Pollmann2010}, 
which is a signature of the underlying SPT phase.   
Figure \ref{fig:es}(a) shows the correct even-fold degeneracy in the low-lying part of the entanglement spectrum,  
which further confirms the presence of the SH phase. 
A remark is in order about the interpretation of ES shown in Fig.~\ref{fig:es}. 
Since our ES are obtained for the fermionic model \eqref{pbandmodelbis},    
some of the higher-lying levels belong to the ``fermionic sector'' of the spectrum and 
may not exhibit the structure expected in bosonic SPT phases, as is demonstrated in, e.g., Refs.~\cite{Hasebe-T-13,Pollmann2015}. 
To resolve this, we separate the bosonic sector (shown by red circles) from the fermionic one (black circles) in Fig.~\ref{fig:es}. 
The degeneracy structure of the bosonic sector now perfectly agrees with what we expect for the corresponding SPT phases.   
In view of the recent developments in entanglement measurements in cold-atom settings \cite{Islam-et-al-entanglement-15}, 
our proposal would make precise characterization of SPT phases possible in experiments. 
In order to show that the SU($N$) SPT phases found above are not restricted to the strong-coupling regime, we plot their extent as a function of $U$ along the physical line $U/V=10$ in Fig.~\ref{fig:phasediag_DMRG} at fixed $t_\perp=t(=1)$.   
These phases occur in the large-$U$ regime and, for weaker interactions, quantum phase transitions are expected
towards fully gapped trivial or dimerized phases which break the translation symmetry spontaneously.  

\emph{Summary and experimental prospects --}
We have introduced a simple one-dimensional microscopic model to describe alkali or alkaline-earth ultracold fermionic atoms loaded into a bichromatic optical lattice. Using analytical and numerical insight, we have shown how SU($N$) SPT phases can emerge in a large range of parameters. This provides a physical route to realize the SH phase ($N=2$),  its generalization for even $N$, as well as  the $\chi$H phase with $N=3$ which breaks spontaneously the inversion symmetry. The experimental realization of the SH phase with $N=2$
may be obtained using the two lowest hyperfine states of ${}^6\mathrm{Li}$.
In this case, the ratio $U/t$ may be tuned using the broad
Feshbach resonance involving these two states \cite{zurn_PRL2013}.
Furthermore, detection resolved in both density and spin is possible
by combining a Fermi-gas microscope with Stern-Gerlach techniques
as done in Ref.~\cite{boll_Science2016} or by ejecting unwanted spin
states using resonant pulses as in Ref.~\cite{parsons_Science2016}. 
The typical temperature scale of recent experiments with ${}^6\mathrm{Li}$ atoms is $T \simeq (0.5-0.8) 4 t^2/U$
\cite{boll_Science2016}. Interestingly enough,  this temperature scale is of the same order of magnitude as the 
gap of the SH phase \cite{DMRG}: $\Delta_{\rm SH} \simeq 0.41 J \simeq  0.8 t^2/U$ obtained
in the large-$U$ limit.
As was recently shown numerically in Ref.~\cite{Becker2017}, the main characteristics of the
  thermal spectral functions of the SH phase with localized edge states are still visible at finite size for $T \simeq \Delta_{\rm SH}$,
a temperature scale which is within the reach of forthcoming experiments. 
Larger values of $N$ are experimentally accessible using
fermionic alkaline-earth or ytterbium atoms.
Using typical experimental values for ${}^{173}\mathrm{Yb}$ which
corresponds to the case of $N=6$ 
(scattering length $a_g=10.55\,\mathrm{nm}$ \cite{kitagawa:PRA2008}
and lattice spacing $\pi/k\approx 400\,\mathrm{nm}$ \cite{hofrichter:PRX2016}),
we find  $U/V \sim 10$. 
Spin-resolved measurements may be performed on these systems using
optical Stern--Gerlach techniques \cite{stellmer_PRA2011}.
In the light of the recent experimental achievements with cold fermionic quantum gases,
we expect the SPT phases discussed in this Letter
to be observed in the near future.

\begin{acknowledgments}
The authors are very grateful to V. Bois for his collaboration at the early stage of this work.
We would like to thank G. Salomon for important discussions.
The authors (SC, PL, and KT) are grateful to CNRS (France) for financial support (PICS grant).  
One of the authors (KT) is supported in part by JSPS KAKENHI Grant No.~15K05211 and No.~JP15H05855.  
This work was performed using HPC resources from GENCI (Grant No. x2016050225 and No. A0010500225) and CALMIP.
Last, the authors thank the program ``Exotic states of matter with SU($N$)-symmetry (YITP-T-16-03)'' held at Yukawa Institute for 
Theoretical Physics where early stage of this work has been carried out.
\end{acknowledgments}


\pagebreak
\widetext
\begin{center}
\textbf{\large Supplemental Materials: ``Haldane'' phases
with ultracold fermionic atoms in double--well optical lattices}
\end{center}
\setcounter{equation}{0}
\setcounter{figure}{0}
\setcounter{table}{0}
\setcounter{page}{1}
\makeatletter
\renewcommand{\theequation}{S\arabic{equation}}
\renewcommand{\thefigure}{S\arabic{figure}}
\renewcommand{\bibnumfmt}[1]{[S#1]}
\renewcommand{\citenumfont}[1]{S#1}

\makeatother

\section{Strong-coupling expansion for SU(3)}
In this section, we consider the strong-coupling limit of SU($N$) cold fermions confined in a double-well optical lattice 
described by the following Hubbard-like Hamiltonian:
\begin{equation}
\begin{split}
{\cal H} =& -t \sum_{i}\sum_{\ell=1}^{2} \sum_{\alpha=1}^{N} \left( 
c_{\ell \alpha,i+1}^\dagger c_{\ell \alpha,i}+ \text{H.c.} \right) 
- t_\perp \sum_{i} \sum_{\alpha=1}^{N} \left( c_{1\alpha,i}^\dagger c_{2 \alpha,i}+ \text{H.c.} \right) 
- \mu \sum_{i} n_i   \\
& + \frac{U}{2}\sum_{i}\sum_{\ell=1}^{2}\sum_{\alpha \neq \beta}n_{\ell  \alpha, i}n_{\ell  \beta, i}  \\
&+ V \sum_{i}\sum_{\alpha \neq \beta} \biggl\{ n_{1 \alpha,i}n_{2 \beta,i}
 +c_{1 \alpha,i}^\dagger c_{2 \beta,i}^\dagger c_{1 \beta,i}c_{2 \alpha,i} 
 + \frac{1}{2}\left(c_{1 \alpha,i}^\dagger c_{1 \beta,i}^\dagger c_{2 \beta,i}c_{2 \alpha,i} + \text{H.c.} \right) \biggr\} \; .
\end{split}
  \label{eq:ham00} 
\end{equation}
Dropping the inter-chain interactions (i.e., $V=0$), we recover the SU($N$) Hubbard ladder.  
The atomic limit ($t=0$) of the above Hamiltonian is most conveniently described 
using the antisymmetric (anti-bonding) and symmetric (bonding) combinations of the $c$-fermions introduced in the Letter:
\begin{equation}
d_{1\sigma,i}=\frac{1}{\sqrt{2}}\left(c_{1 \sigma,i}-c_{2 \sigma,i}\right)  \; , \;\; 
d_{2 \sigma,i}=\frac{1}{\sqrt{2}}\left(c_{1 \sigma,i}+c_{2 \sigma,i}\right) \; ,
\end{equation}
in terms of which the atomic-limit Hamiltonian reads as [$t=0$ limit of Eq.~(4) of the Letter]:
\begin{equation}
\begin{split}
& \mathcal{H}_{0} = \sum_{i} \mathcal{H}_{\text{on-site}}(i)  \\
&  \mathcal{H}_{\text{on-site}}(i)  =  - \left( \mu + \frac{U+V}{2} \right)   n_i + 2 t_\perp  T_i^z  +  \frac{U + V}{4}  n_i^2 
+ 2 V  (T_i^z)^2  +  (U-V) (T_i^x)^2 \; .
\end{split}
  \label{pbandmodelbissupp}
\end{equation}
The orbital pseudo-spin $\mathbf{T}$ is defined with respect to the $d$-fermions: 
$T_i^a = \frac{1}{2} \sum_{m,n,\alpha} d_{m \alpha,\,i}^\dag \sigma^a_{m n} d_{n \alpha,\,i} $.  Note that the hopping between the two wells 
($\ell=1,2$) now translates to the number difference: $t_{\perp} (n^{(d)}_{1,i}-n^{(d)}_{2,i})$.      
The spectrum of $\mathcal{H}_{\text{on-site}}$ \eqref{pbandmodelbissupp} is labeled by various quantum numbers, i.e.,  
(i) the total number of particle $n_i$, 
(ii) the orbital pseudo-spin squared $\mathbf{T}_i^2=T(T+1)$ ($T^{z}$ is not a good quantum number in general) as well as 
(iii) the SU($N$) irreducible representations which are most conveniently specified by Young diagrams 
with at most two columns \cite{Itzykson-N-66,Georgi-book-99}.   
Although the on-site part of the Hamiltonian does not contain SU($N$)-dependent interactions, 
the optimal SU($N$) representation is selected by the orbital($\mathbf{T}$)-dependent part through 
the Fermi statistics (see, e.g., Appendix A of Ref.~\cite{Bois-C-L-M-T-15supp}).  
The condition of half filling is imposed by setting 
\begin{equation}
\mu=\frac{N}{2}(U+V)
\end{equation}
for which the spectrum exhibits the particle-hole symmetry: $ n \leftrightarrow 2N-n$. 
To ease the notations, we will drop the site index for the on-site limit spectrum.

\subsection{Atomic-limit spectrum}
The atomic-limit Hamiltonian \eqref{pbandmodelbissupp} commutes with the SU($N$) generators and the orbital pseudo-spin 
$\mathbf{T}$, we can diagonalize it for given SU($N$) representation and $\mathbf{T}$.  
Due to the fermionic statistics, only special combinations of SU($N$) representations and 
$\mathbf{T}$ appear for a given local fermion number $n$ ($0 \leq n \leq 2N$) \cite{Bois-C-L-M-T-15supp}:
 \begin{subequations}
 \begin{align}
 & \yng(1) \sim (\underbrace{\yng(1)}_{\text{SU}(N)},\underbrace{\yng(1)}_{\text{SU(2)}}) 
 \quad (n =1) \\
 & \yng(1,1) \sim \left(\yng(2),\bullet \right) \oplus \left(\yng(1,1), \yng(2) \right)  \quad (n =2) \\
 & \yng(1,1,1) \sim \left( \yng(2,1), \yng(1) \right) \oplus \left(\yng(1,1,1),\yng(3) \right)  \quad (n =3) \\
 & \yng(1,1,1,1) \sim \left( \yng(2,2), \bullet \right) \oplus \left( \yng(2,1,1),\yng(2) \right) 
 \oplus \left(\yng(1,1,1,1), \yng(4) \right)  \quad (n=4) 
 \label{eqn:decomp-SU2-SUN-4} \\
 & \yng(1,1,1,1,1) \sim \left( \yng(2,2,1), \yng(1) \right) \oplus \left( \yng(2,1,1,1),\yng(3) \right) 
 \oplus \left(\yng(1,1,1,1,1), \yng(5) \right)  \quad (n =5) \\
 & \qquad \qquad \qquad \vdots \notag \\
 & \text{\scriptsize $2N$} \left\{ 
\yng(1,1,1,1,1)
\right. 
\sim \left( \bullet, \bullet \right)   \quad (n =2N) 
\end{align}
\label{eqn:decomp-SU2-SUN}
\end{subequations}
As the Hamiltonian \eqref{pbandmodelbissupp} does not depend on SU($N$), given $n$, we just diagonalize \eqref{pbandmodelbissupp} 
for all allowed $T$.  

\begin{figure}[!ht]
\centering
\includegraphics[scale=0.8]{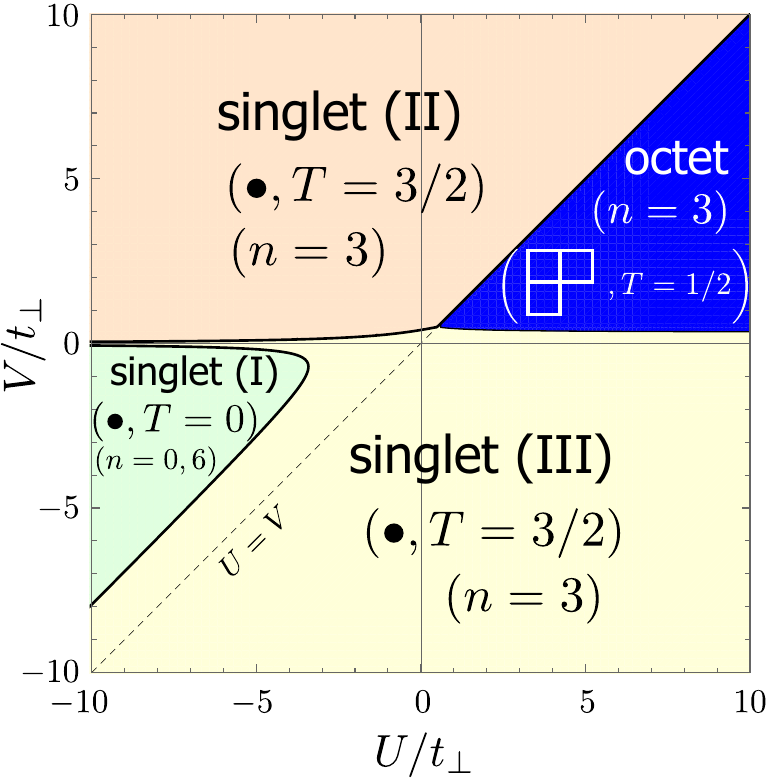}
\caption{Atomic-limit phase diagram of $\mathcal{H}_{\text{on-site}}$ for $N=3$.   
There are four phases: 
(i) SU(3) octet phase with $n=3$, $\mathbf{8}$ (adjoint), and $(T=1/2,T^{z}=-1/2)$ [blue; second of Eq.~\eqref{eqn:N3-adjoint-T1o2}], 
(ii) SU(3)-singlet phase (I) with $n=0$ or $6$, $\bullet$, and $T=0$ [green; Eq.~\eqref{eqn:double-well-energy-adjoint}],  
(iii) SU(3)-singlet phase (II) with $n=3$, $\bullet$, and $T=3/2$ [pale orange; first of Eq.~\eqref{eqn:N3-quartet}], 
and (iv) SU(3)-singlet phase (III) with $n=3$, $\bullet$, and $T=3/2$ [pale yellow; third of Eq.~\eqref{eqn:N3-quartet}].  
The SU(3)-singlet phase (I) suggests that a period-2 charge-density wave might be stabilized there.  
Chemical potential is fixed to $\mu=3(U + V)/2$.
\label{fig:CouplagefortN3tp1}}
\end{figure}

When $N=3$, the fermion number can take $n=0,1,\ldots, 6\,(=2N)$.  
The energy-spectrum for the fermion number $n$ is given by:
\begin{subequations}
\begin{align}
\label{eqn:double-well-energy-adjoint}
& E_{n=0,6}(\bullet,T=0) = 0  \\
& E_{n=1,5}((\mathbf{3},\bar{\mathbf{3}}),T=1/2) = 
\begin{cases}
t_\perp-\left(U+V\right) & (T^z=+1/2; \times 3 ) \\
-t_\perp-\left(U+V\right) & (T^z=-1/2; \times 3)
\end{cases} \\
& E_{n=2,4}((\mathbf{6},\bar{\mathbf{6}}),T=0) = -2\left(U+V\right) \; (\times 6) \\
& E_{n=2,4}((\bar{\mathbf{3}},\mathbf{3}),T=1) = 
\begin{cases}
-\frac{3}{2}U-\frac{1}{2}V-\sqrt{4 t_\perp^2+\left(\frac{U-V}{2}\right)^2} & (\times 3) \\
-\frac{3}{2}U-\frac{1}{2}V+\sqrt{4 t_\perp^2+\left(\frac{U-V}{2}\right)^2} & (\times 3) \\
-U-3V & (T^z=0;  \times 3) 
\end{cases}  \\
& E_{n=3}(\mathbf{8},T=1/2) = 
\begin{cases}
t_\perp -2(U+V) & T^z=+1/2 \;\; (\times 8) \\
-t_\perp -2(U+V) & T^z=-1/2 \;\; (\times 8) 
\end{cases} 
\label{eqn:N3-adjoint-T1o2}
\\
& E_{n=3}(\bullet,T=3/2) = 
\begin{cases}
t_\perp - U - V - \sqrt{4 t_\perp^2 - 2 t_\perp U + U^2 + 10 t_\perp V - 4 U V + 7 V^2} & \\
t_\perp - U - V + \sqrt{4 t_\perp^2 - 2 t_\perp U + U^2 + 10 t_\perp V - 4 U V + 7 V^2} & \\
-t_\perp - U - V - \sqrt{4 t_\perp^2 + 2 t_\perp U + U^2 - 10 t_\perp V - 4 U V + 7 V^2} & \\
-t_\perp - U - V + \sqrt{4 t_\perp^2 + 2 t_\perp U + U^2 - 10 t_\perp V - 4 U V + 7 V^2} & ,
\end{cases}
\label{eqn:N3-quartet}
\end{align}
\end{subequations}
where the irreducible representations of the SU(3) group are labeled by their dimension: 
\begin{equation}
\begin{split}
& \mathbf{3} \; \leftrightarrow \; \yng(1) \; , \;\; 
\bar{\mathbf{3}} \; \leftrightarrow \;  \yng(1,1) \; ,  \\
& \mathbf{6} \;  \leftrightarrow \; \yng(2) \; , \; 
\bar{\mathbf{6}} \;   \leftrightarrow \;  \yng(2,2) \\
& \mathbf{8} \; \leftrightarrow \; \yng(2,1) \; . 
\end{split}
\end{equation}
The lowest of these energies for different parameters ($U,V$) are displayed in Fig. (\ref{fig:CouplagefortN3tp1}).
\subsection{Effective Hamiltonian for octet phase}
Now let us consider the octet phase (shown by blue in Fig.~\ref{fig:CouplagefortN3tp1}) 
where an SU(3) ``magnetic moment'' in the $\mathbf{8}$ (adjoint) representation 
is formed at each rung (i.e., double well) and the orbital pseudo-spin $\mathbf{T}$ is quenched to $T=1/2,T^z=-1/2$.  
In this phase, the low-energy effective Hamiltonian is expected to be SU(3)-invariant and written only in terms of 
the SU(3) ``spins'' in $\mathbf{8}$.   
First, we restrict the form of possible interactions by symmetry consideration.  
From the Clebsch-Gordan decomposition 
\begin{equation}
\begin{split}
& \yng(2,1) \otimes \yng(2,1) \quad \simeq \quad 
\underbrace{\yng(4,2)}_{\mathbf{27}} \; \oplus \; \underbrace{\yng(3)}_{\mathbf{10}} 
\; \oplus \; \underbrace{\yng(3,3)}_{\overline{\mathbf{10}}} \; \oplus \; \underbrace{\yng(2,1)}_{\mathbf{8}_{\text{S}}}  
\; \oplus \; \underbrace{\yng(2,1)}_{\mathbf{8}_{\text{A}}} \; \oplus \; \bullet \\
& (\mathbf{8} \times \mathbf{8} = \mathbf{27} \oplus \mathbf{10} \oplus \overline{\mathbf{10}} 
\oplus \mathbf{8}_{\text{S}} \oplus \mathbf{8}_{\text{A}} \oplus \mathbf{1} ) \; ,
\end{split}
\label{eqn:decomp-8x8}
\end{equation}
(the subscripts ``S'' and ``A'' label the two $\mathbf{8}$ representations)  
one sees (by the Schur's lemma) 
that any SU(3)-invariant (two-site) interactions can be completely parametrized by six independent coefficients 
corresponding to the six irreducible representations appearing on the right-hand side:
\begin{equation}
 a_{\mathbf{27}} P_{\mathbf{27}} + a_{\mathbf{10}} P_{\mathbf{10}} 
+ a_{\overline{\mathbf{10}}} P_{\overline{\mathbf{10}}} 
+ a_{\mathbf{8}_{\text{S}}} P_{\mathbf{8}_{\text{S}}} + a_{\mathbf{8}_{\text{A}}} P_{\mathbf{8}_{\text{A}}} 
+ a_{\mathbf{1}} P_{\mathbf{1}}  
\label{eqn:SU(3)-Ham-general}
\end{equation}
with $P_{R}$ being the projection operator onto the irreducible represenation $R$ and $a_{R}$ the corresponding 
real coefficient.   
In the case of SU(2), the projection operators are uniquely expressed in terms of polynomials of the quadratic Casimir $\mathcal{C}_{2}$ 
(or, $\mathbf{S}_{1}{\cdot}\mathbf{S}_{2}$).  On the other hand, in SU($N$) ($N\geq 3$), higher-order Casimirs 
and other operators may also be necessary to recast the general form \eqref{eqn:SU(3)-Ham-general} into the ``spin'' Hamiltonian.  
Specifically, for a pair of SU(3) spins $\mathcal{S}_{1}^{A}$, $\mathcal{S}_{2}^{B}$ 
in $\mathbf{8}$, the most general SU(3)-invariant two-site interaction may be written as:
\begin{equation}
{\cal H}^{(\text{octet})}_{\text{eff}} = \alpha \mathbf{1}  
+ J_1  \sum_A \mathcal{S}^A_1 \mathcal{S}^A_2 
+J_2\left(\sum_A \mathcal{S}^A_1 \mathcal{S}^A_2\right)^2 
+J_3\left(\sum_A \mathcal{S}^A_1 \mathcal{S}^A_2\right)^3 
+ J_4 \mathcal{C}_{3}(\mathcal{S}_{1},\mathcal{S}_{2}) 
+p \mathcal{P}_{\mathbf{8},\mathbf{8}} ,
\label{hameffN3}
\end{equation}
where $\mathcal{P}_{\mathbf{8},\mathbf{8}}$ is the permutation operators of the neighboring sites 1 and 2, and 
$\mathcal{C}_{3}$ is the cubic Casimir made of $\mathcal{S}_{1,2}^{A}$.    
The ``exchange interaction'' $\sum_{A=1}^{8} \mathcal{S}^A_1 \mathcal{S}^A_2$ is directly related to the quadratic Casimir as:
\begin{equation}
\begin{split}
\sum_{A=1}^{8} \mathcal{S}^A_1 \mathcal{S}^A_2
& =\sum_{A} \frac{1}{2}\left\{ 
\left(\mathcal{S}^A_1 +\mathcal{S}^A_2\right)^2 
-\left(\mathcal{S}^A_1\right)^2-\left(\mathcal{S}^A_2\right)^2
\right\}  \\
& =  \frac{1}{2}\left\{
\mathcal{C}_{2}(R) 
- 2 \mathcal{C}_{2}(\mathbf{8}) 
\right\}  ,
\end{split}
\end{equation}
which enables us to use $\sum_{A=1}^{8} \mathcal{S}^A_1 \mathcal{S}^A_2$ instead of the full quadratic Casimir operator.  
The reason for the necessity of $\mathcal{P}_{\mathbf{8},\mathbf{8}}$ is that the two adjoint representations $\mathbf{8}$ 
share the same set of the Casimirs $(\mathcal{C}_{2},\mathcal{C}_{3})$ (see Table \ref{tab:Casimir-6irreps}) 
and are distinguished only by $\mathcal{P}_{\mathbf{8},\mathbf{8}}$.   
As $\mathcal{C}_{3}$ is odd under the conjugation $R \to \overline{R}$ (which, in the fermion language, 
translates to the particle-hole transformation), $J_{4}=0$ at half-filling.  

\begin{figure}[h]
\begin{center}
\includegraphics[scale=0.6]{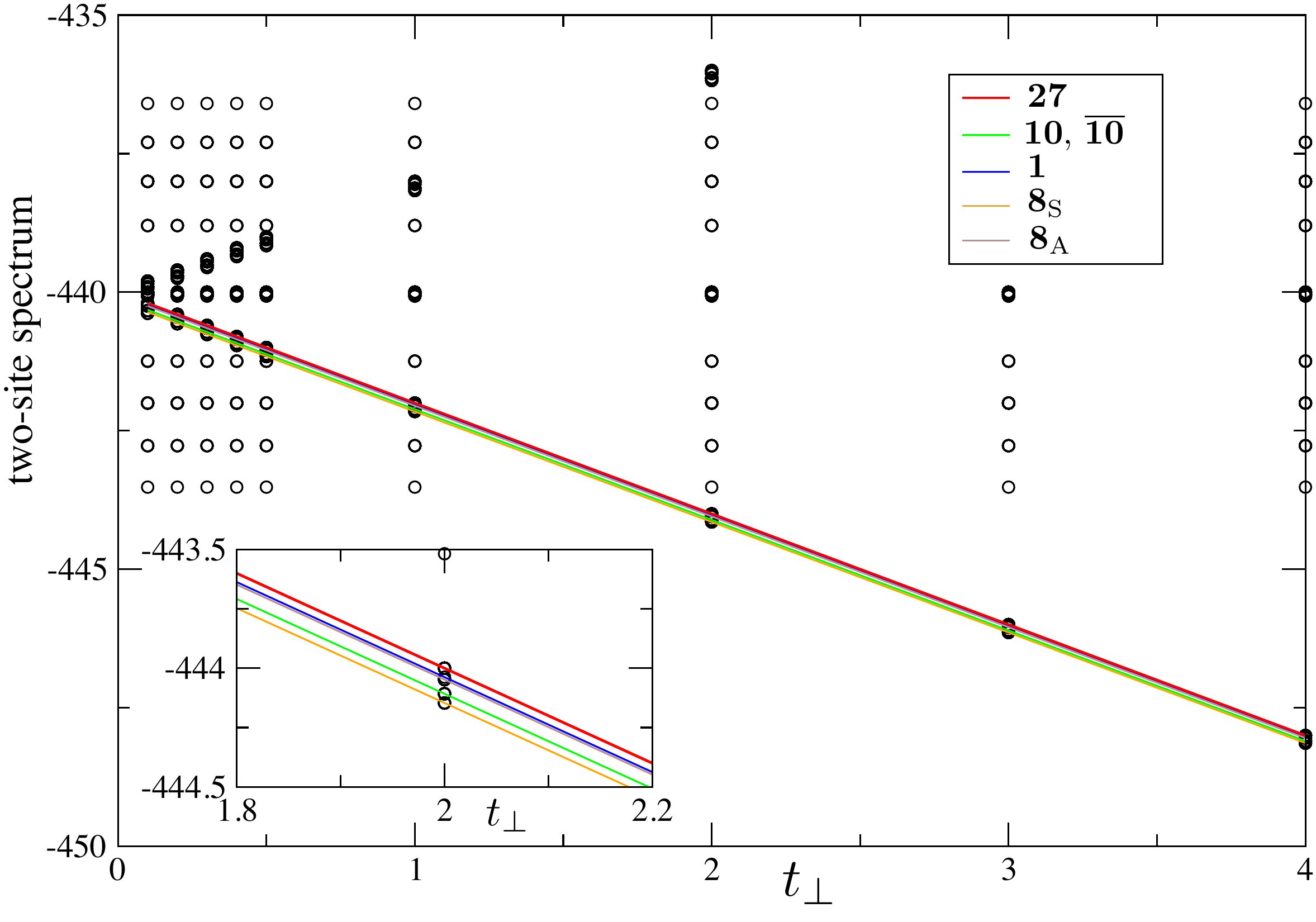}
\end{center}
\caption{Exact two-site spectrum for $U=100$, $V=10$ and unit $t$ versus $t_\perp$. The spectrum \eqref{eqn:effective-spectrum} predicted by the effective Hamiltonian \eqref{hameffN3} is drawn in color lines. A zoom on the ground states around $t_\perp=2$ is provided on the bottom left corner. The effective Hamiltonian description breaks down for $t_\perp<1.7t$.}
\label{fig:ergs_vs_tperp.U.100.V.10}
\end{figure}

\begin{figure}[!ht]
\centering
\includegraphics[width=0.8\columnwidth,clip]{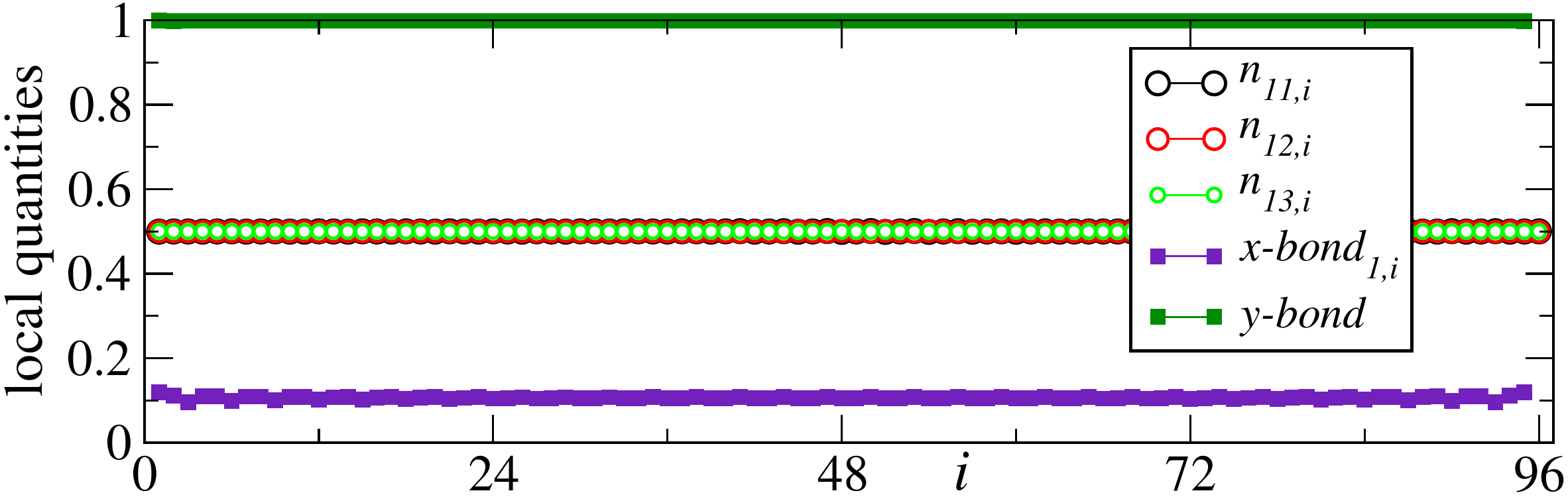}
\caption{Local quantities for the $N=3$ model obtained by DMRG simulations for $L=96$, $(U,V)=(100,10)$ and $t_\perp/t=10$.}
\label{fig:QPT_N3c}
\end{figure}

Second-order processes in $t$ give an effective interaction between a pair of SU(3) spins in $\mathbf{8}$, 
whose coupling constants $\{\alpha, J_1,J_2,J_3,p\}$ are given in terms of $(t,t_{\perp},U,V)$ as:
\begin{subequations}
\begin{align}
& \alpha = \frac{1}{6}\left(\frac{A}{6}-\frac{B}{2}\right) \\
& J_1 = \frac{A}{9}+\frac{2B}{9} \, , \; \; 
J_2 = - \frac{10}{27}\left(A - B \right) \, , \; \; 
J_3 = - \frac{2}{27}\left(A - B \right) \\ 
& p = \frac{1}{6}\left(A - B \right) ,
\end{align}
where 
\begin{equation}
\begin{split}
A & \equiv  \frac{t^2\left(8 t_\perp +3U+ V\right)}{U\left( U+V\right)+t_\perp \left(3U+V\right)}\\
B & \equiv   \frac{3t^2\left(8 t_\perp +U+ 3V\right)}{V\left( U+V\right)+t_\perp \left(U+3V\right)} \; .
\end{split}
\end{equation}
\end{subequations}
The spectrum of the two-site effective Hamiltonian ${\cal H}_{\text{eff}}$ \eqref{hameffN3} reads as:
\begin{subequations}
\begin{align}
&  E_0 \quad (\times 27; \;\; \mathbf{27}) \\
& E_0 - \frac{B}{3} \quad (\times 20; \;\; \mathbf{10}, \overline{\mathbf{10}}) \\
& E_0 - \frac{4A}{3}  \quad (\times 1; \;\; \mathbf{1}) \\
& E_0 - \frac{5(A+B)}{12}    \quad (\times 8; \; \mathbf{8}_{\text{S}}) \, , \;\; 
 E_0 - \frac{3A}{4}-\frac{B}{12}  \quad   (\times 8; \; \mathbf{8}_{\text{A}}) .
\end{align}
\label{eqn:effective-spectrum}
\end{subequations}
with $E_{0}=2E_{n=3}(\mathbf{8},T=1/2) = -2t_\perp -4(U+V)$.  
For values of parameters corresponding to the octet phase of Fig. \ref{fig:CouplagefortN3tp1},  we have $J_1 \sim J_2$ 
and we must retain the biquadratic term $\left(\sum_A \mathcal{S}^A_1 \mathcal{S}^A_2\right)^2$, 
and hence nothing more can be said about the physics described without a direct numerical investigation of the effective Hamiltonian \eqref{hameffN3}. Furthermore a numerical derivation of the full two-site spectrum shows that the effective Hamiltonian description is valid for $t_\perp \gtrsim  t$  ($t_\perp > 1.7 t$ in the case of $U=100$, $V=10$ for unit $t$ according to Fig.~\ref{fig:ergs_vs_tperp.U.100.V.10}). Given the parameters dependences of $J_1, J_2, J_3$ and $p$, their values stay qualitatively the same on all the octet region of Fig.~\ref{fig:CouplagefortN3tp1}. Hence Fig.~\ref{fig:QPT_N3c} obtained by DMRG for $t_{\perp} = 10$, $t=1$, $U=100$, and $V=10$ provides also a valid depiction of the phase described by the effective Hamiltonian \eqref{hameffN3} for realistic parameters. One thus concludes that for $N=3$ a featureless fully gapped phase is stabilized without any edge states. However, this description breaks down when $t_\perp \lesssim  t$, the typical regime where a SPT phase can emerge, as shown below, where we have to use the other approach presented in the Letter.  

\begin{table}[!ht]
\begin{center}
\begin{tabular}{lcccccc}
\hline\hline
irreps. & ${\tiny \yng(4,2)}$ & ${\tiny \yng(3)}$ & ${\tiny \yng(3,3)}$ & ${\tiny \yng(2,1)}$ & ${\tiny \yng(2,1)}$ & $\bullet$ \\
\hline
dimensions & $\mathbf{27}$ & $\mathbf{10}$ & $\overline{\mathbf{10}}$ & $\mathbf{8}_{\text{S}}$ & $\mathbf{8}_{\text{A}}$ & $\mathbf{1}$ \\
quadratic Casimir $\mathcal{C}_{2}$ & 8 & 6 & 6 & 3 & 3 & 0 \\ 
cubic Casimir $\mathcal{C}_{3}$ & 0 & 9 & $-9$ & 0 & 0 & 0 \\ 
symmetry ($\mathcal{P}_{\mathbf{8},\mathbf{8}}$) & $1$ & $-1$ & $-1$ & $1$ & $-1$ & $1$ \\
\hline\hline
\end{tabular}
\caption{\label{tab:Casimir-6irreps} Quadratic and cubic Casimirs for the six irreducible representations 
appearing in the decomposition of $\mathbf{8}\otimes \mathbf{8}$ [eq.\eqref{eqn:decomp-8x8}].  
}
\end{center}
\end{table}

\section{Numerical results}
\subsection{Edge states in $N=3$ model}
\begin{figure}[h]
\begin{center}
\includegraphics[scale=0.6]{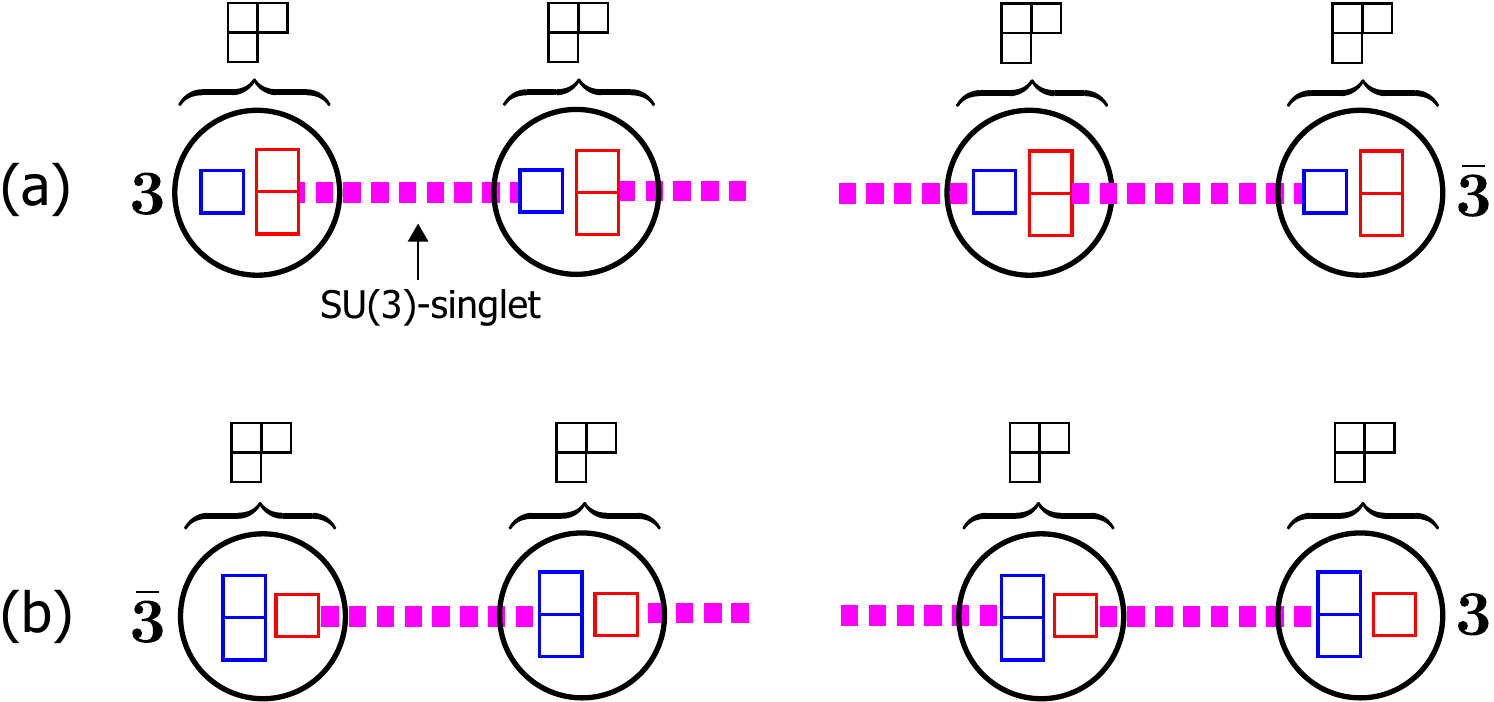}
\end{center}
\caption{Two SU(3) VBS states that break reflection symmetry spontaneously.  
They are distinguished by the edge states: in (a), $\mathbf{3}$ and $\bar{\mathbf{3}}$ respectively appear 
at the left and right edges, while, in (b), $\mathbf{3}$ and $\bar{\mathbf{3}}$ are interchanged.  
We call these VBS states $(\mathbf{3},\bar{\mathbf{3}})$ [(a)] and $(\bar{\mathbf{3}},\mathbf{3})$ [(b)].  
\label{fig:SU3-8rep-MPS}}
\end{figure}

As is well-known, the sharpest characterization of (bosonic) SPT phases is obtained by analyzing 
the projective representation appearing at the edges (whether physical or virtual).  
Nevertheless, for practical purposes, the observation of the {\em physical} edge states still provides us with 
a useful method of accessing the underlying topological properties (especially when we consider bosonic SPT phases 
realized in {\em fermionic} systems).   
To demonstrate how this strategy works, let us consider the two SU(3) valence-bond solid (VBS) states \cite{Affleck-K-L-T-88} and calculate the expectation values of the two Cartan generators $G^{3}_{r}$ and $G^{8}_{r}$ 
in the 8-dimensional adjoint representation ($\mathbf{8}$) at site $r$ 
(the SU(3) generators are normalized as $\text{Tr} G^{A}G^{B}=6 \delta^{AB}$).   
The two VBS states break the reflection symmetry spontaneously 
and are known \cite{Morimoto-U-M-F-14} to belong to the two different SPT phases predicted by 
group cohomology \cite{Duivenvoorden-Q-13supp} (see Fig.~\ref{fig:SU3-8rep-MPS}).  

\begin{figure}[h]
\begin{center}
\includegraphics[scale=0.8]{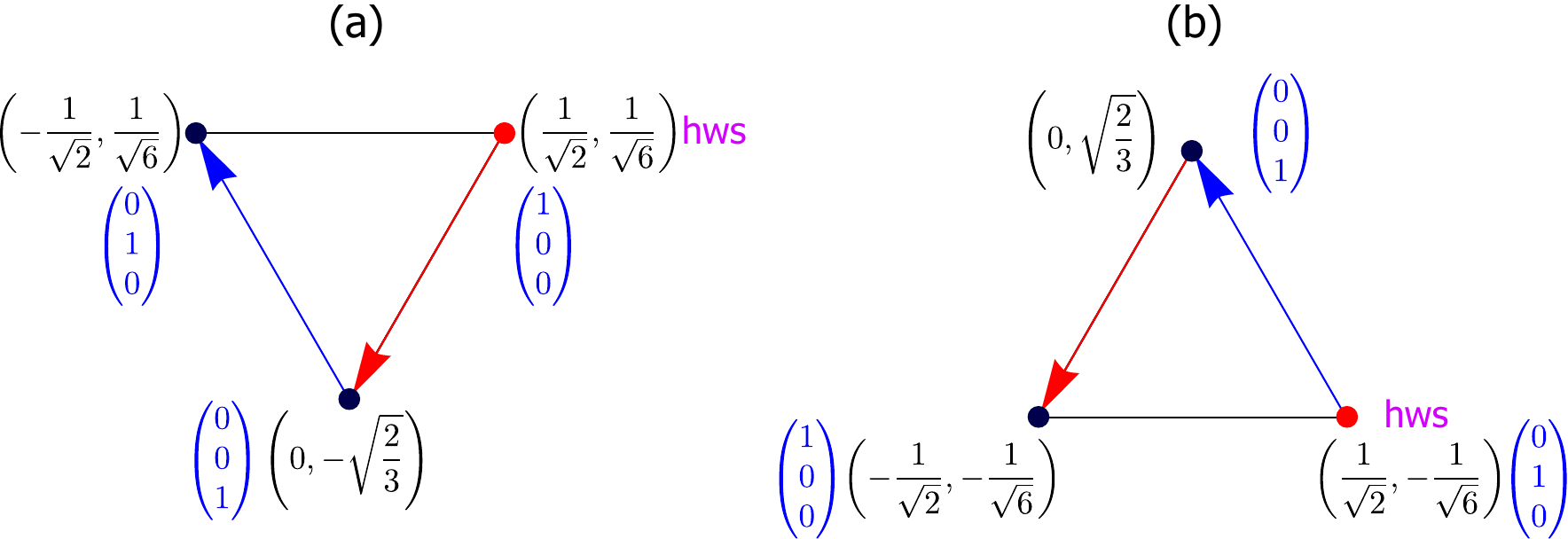}
\end{center}
\caption{Weights of 3 representation ($\mathbf{3}$) and its conjugate ($\bar{\mathbf{3}}$).   
Red (blue) arrows denote the action of simple root $-\boldsymbol{\alpha}_{1}$ ($-\boldsymbol{\alpha}_{2}$).  
\label{fig:weights-SU3-3-3bar-rep}}
\end{figure}

We begin with the $(\mathbf{3},\bar{\mathbf{3}})$ VBS state shown in Fig.~\ref{fig:SU3-8rep-MPS}(a).  
Using the matrix-product state formalism, it is straightforward to calculate the local expectation values of $G^{3}_{r}$ and $G^{8}_{r}$ 
for a semi-infinite system (we have chosen a semi-infinite system just to suppress the effects from the other edge): 
\begin{equation}
\left( \VEV{G^{3}_{r}}, \VEV{G^{8}_{r}} \right) = 
\begin{cases}
\left(  -\frac{9}{\sqrt{2}} \left(-\frac{1}{8} \right)^r ,  -\frac{3\sqrt{3}}{\sqrt{2}} \left(-\frac{1}{8} \right)^r  \right) & \text{for L-edge state 1} \\
\left(  +\frac{9}{\sqrt{2}} \left(-\frac{1}{8} \right)^r ,  -\frac{3\sqrt{3}}{\sqrt{2}} \left(-\frac{1}{8} \right)^r  \right) & \text{for L-edge state 2} \\
\left( 0, 3\sqrt{6} \left(-\frac{1}{8} \right)^r   \right)  & \text{for L-edge state 3}  \; .
\end{cases}
\end{equation}
Summing up these values, we obtain the edge moment localized around the left edge:
\begin{equation}\label{eq:G3G8L}
\sum_{r=1}^{\infty} 
\left( \VEV{G^{3}_{r}}, \VEV{G^{8}_{r}} \right) = 
\begin{cases}
\left( \frac{1}{\sqrt{2}}, \frac{1}{\sqrt{6}} \right)=\lambda_{\text{\bf 3}}(1) & \text{for L-edge state 1 (heighest weight state (hws) of $\mathbf{3}$)} \\
\left( - \frac{1}{\sqrt{2}}, \frac{1}{\sqrt{6}} \right)=\lambda_{\text{\bf 3}}(2) & \text{for L-edge state 2} \\
\left(0 , -\sqrt{\frac{2}{3}} \right) =\lambda_{\text{\bf 3}}(3) & \text{for L-edge state 3}  
\end{cases}
\end{equation}
which is to be compared with the SU(3) weights of $\mathbf{3}$ shown in Fig.~\ref{fig:weights-SU3-3-3bar-rep} (a).   
A similar calculation leads us to the following set of weights at the right edge:
\begin{equation}\label{eq:G3G8R}
\sum_{r=\text{right edge}} 
\left( \VEV{G^{3}_{r}}, \VEV{G^{8}_{r}} \right) = 
\begin{cases}
\left( -\frac{1}{\sqrt{2}}, -\frac{1}{\sqrt{6}} \right)= \lambda_{\bar{\text{\bf 3}}}(1) & \text{for R-edge state 1} \\
\left(  \frac{1}{\sqrt{2}}, - \frac{1}{\sqrt{6}} \right)= \lambda_{\bar{\text{\bf 3}}}(2) & \text{for R-edge state 2 (hws of $\bar{\mathbf{3}}$)} \\
\left(0 , \sqrt{\frac{2}{3}} \right) = \lambda_{\bar{\text{\bf 3}}}(3) & \text{for R-edge state 3}  \; , 
\end{cases}
\end{equation}
which then implies that $\bar{\mathbf{3}}$ appears at the right edge.  
(For the other topological VBS state $(\bar{\mathbf{3}},\mathbf{3})$, we just obtain the weights with $\mathbf{3}$ and 
$\bar{\mathbf{3}}$ interchanged.) 

In the Letter, we have shown that, in some region of the Mott-insulating phase, the above SU(3) SPT phases are stabilized.  
In the region, we may expect that the following fermion operators reduce to the SU(3) ``spins'' $G_{r}^{A}$ in $\mathbf{8}$:
\begin{equation}
\begin{split}
\widehat{G}_{r}^{A} 
& = \sum_{\alpha,\beta=1}^{3}c_{1\alpha,r}^{\dagger}\left( G^{A}\right)_{\alpha\beta} c_{1\beta,r} 
+ \sum_{\alpha,\beta=1}^{3} c_{2\alpha,r}^{\dagger}\left( G^{A}\right)_{\alpha\beta} c_{2\beta,r}   \\
& = \sum_{\alpha,\beta=1}^{3} d_{1\alpha,r}^{\dagger}\left( G^{A}\right)_{\alpha\beta} d_{1\beta,r} 
+ \sum_{\alpha,\beta=1}^{3} d_{2\alpha,r}^{\dagger}\left( G^{A}\right)_{\alpha\beta} d_{2\beta,r}  \quad (A=3,8) \\
& = 
\begin{cases}
\frac{1}{\sqrt{2}} (n_{r}(1) - n_{r}(2) ) & A=3 \\
\frac{1}{\sqrt{6}} (n_{r}(1) + n_{r}(2) - 2 n_{r}(3)) & A=8  \; ,
\end{cases}
\end{split}
\end{equation}
where $n_{r}(\alpha)= \sum_{\ell=1,2}c_{\ell\alpha,r}^{\dagger}c_{\ell\alpha,r}= \sum_{\ell=1,2}d_{\ell\alpha,r}^{\dagger}d_{\ell\alpha,r}$.  

\begin{figure}[h]
\begin{center}
\includegraphics[scale=0.4]{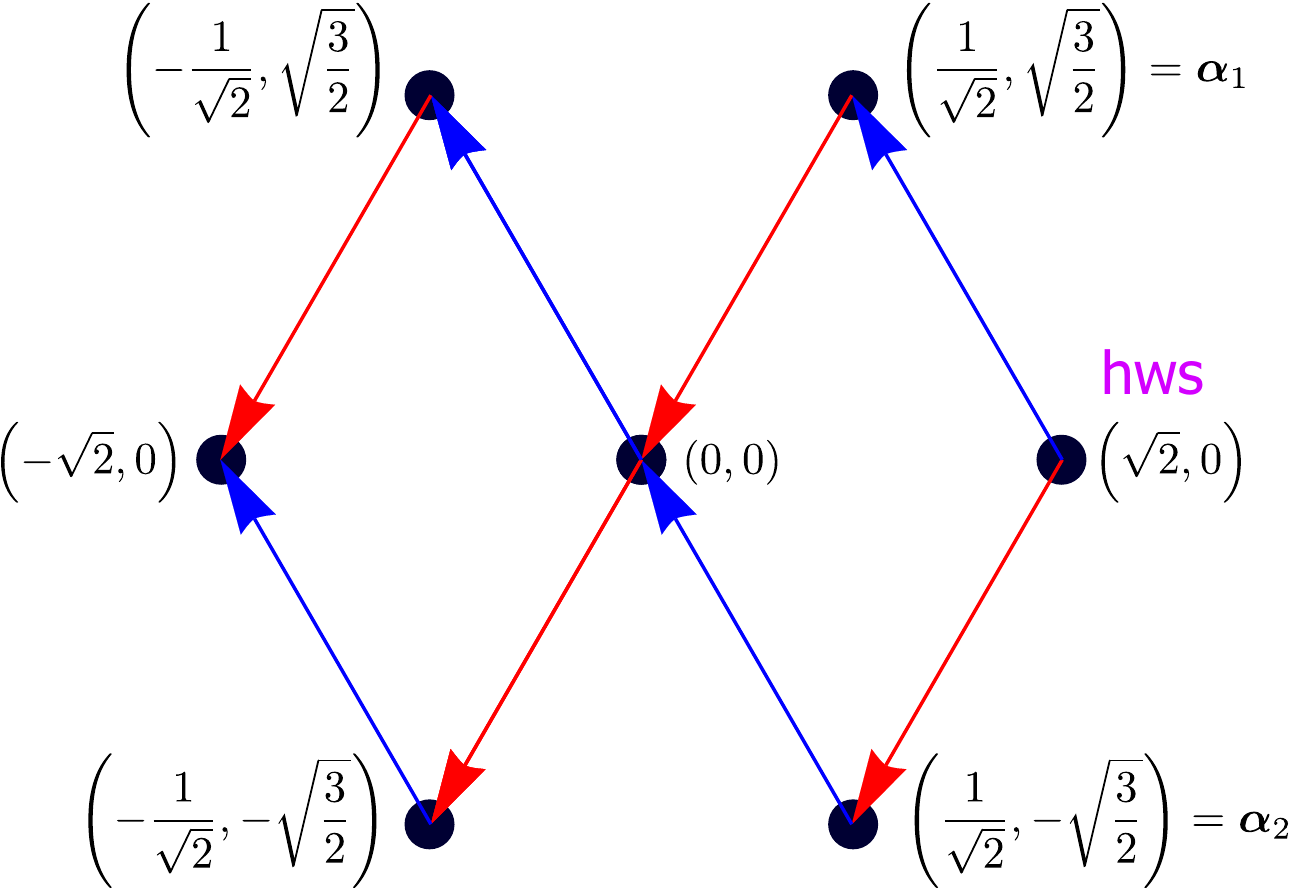}
\end{center}
\caption{Weights 8-dimensional (adjoint) representation.  
The weight $(0,0)$ is doubly degenerate [corresponding to the two Cartan generators of SU(3)].  
Red (blue) arrows denote the action of simple root $-\boldsymbol{\alpha}_{1}$ ($-\boldsymbol{\alpha}_{2}$).  
\label{fig:weights-SU3-8-rep}}
\end{figure}

In the SU(3) SPT phase, we thus expect an overall $2\times 3 \times 3 = 18$ degeneracy of the ground-state corresponding to all possible edge states as well as inversion symmetry. In a numerical DMRG simulation, it is well-known that the algorithm will converge to one of these ground-states randomly (i.e. convergence depends on the sweeping procedure and other details). Note also that since we are implementing U(1) quantum numbers corresponding to the color conservation, left- and right-edge states are related so that $(G^3_\mathrm{tot},G^8_\mathrm{tot})=(0,0)$, which reduces the degeneracy to $2\times 3=6$. In order to characterize a given edge states, we can simply use the local densities $n_{\ell\alpha,i}$ to compute quantities in Eqs.~\eqref{eq:G3G8L}-\eqref{eq:G3G8R}. 
For instance, the data presented in Fig.~2 of the main text for $N=3$ correspond to a left-edge having $\lambda_{\bar{\text{\bf 3}}}(1)$. 

\begin{figure}[!ht]
\centering
\includegraphics[width=0.8\columnwidth,clip]{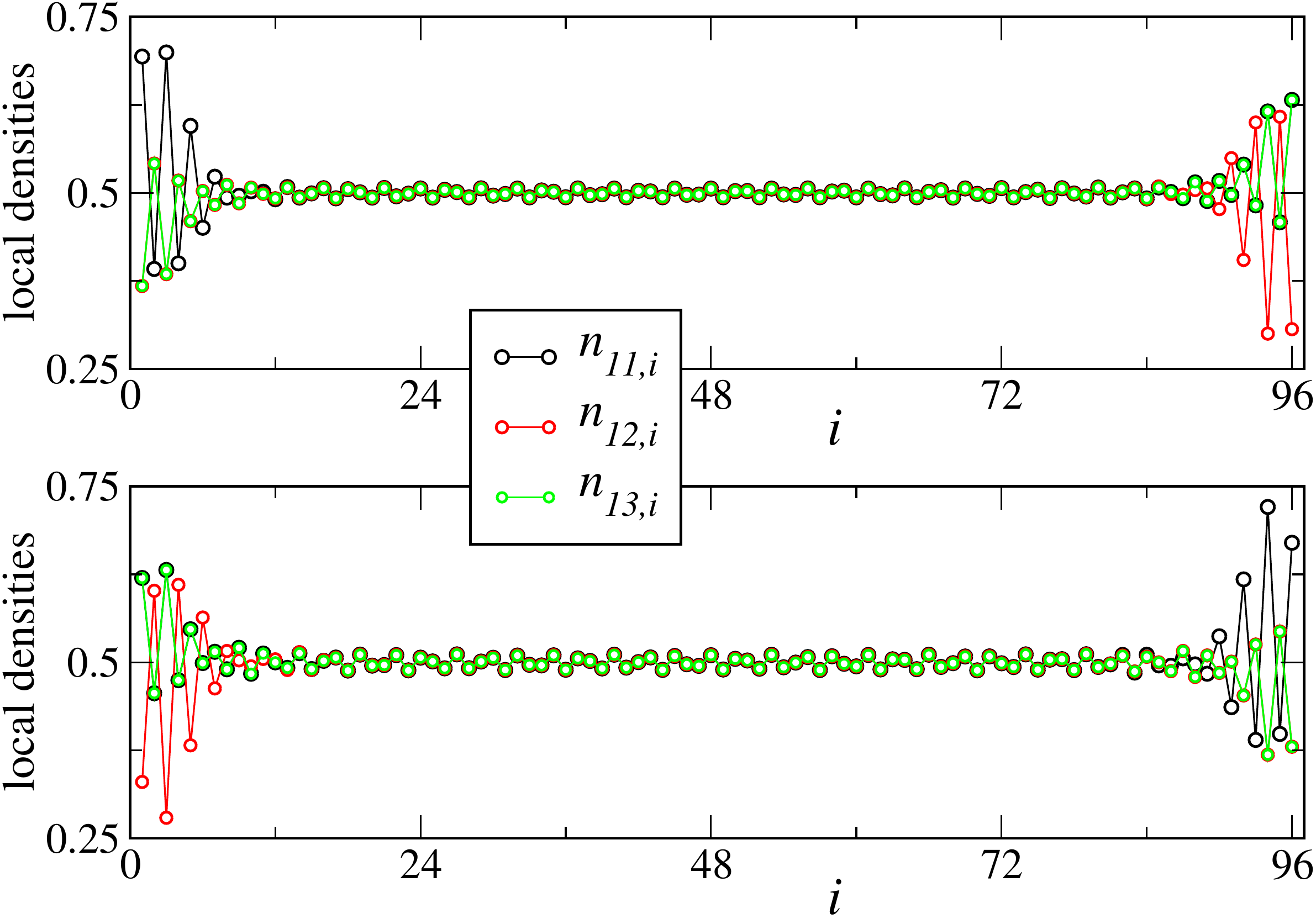}
\caption{Local densities for the $N=3$ model obtained by DMRG simulations for $L=96$, $t=t_\perp=1$ and 
$(U,V)=(20,2)$ (top) or $(U,V)=(100,10)$ (bottom) respectively. We fix the total number of particles per color as 
$(n_1,n_2,n_3)=(L+1,L-1,L)$ and we do observe the two possible edge states (corresponding to the two ground-states in this sector, see text). Top: left-edge has $\lambda_{\bar{\text{\bf 3}}}(1)$ and right-edge 
$\lambda_{\bar{\text{\bf 3}}}(2)$. Bottom:  opposite situation showing the inversion symmetry breaking.}
\label{fig:edgeN3}
\end{figure}

In order to further reduce the degeneracy, we can also work in a ``polarized'' case (by analogy with the spin-1 case for instance) by fixing the total number of particles per color as $(n_1,n_2,n_3)=(L+1,L-1,L)$ so that $(G^3_\mathrm{tot},G^8_\mathrm{tot})=(\sqrt{2},0)$ for which there are only two candidates, namely left-edge having 
$\lambda_{\text{\bf 3}}(1)$  and right-edge $\lambda_{\bar{\text{\bf 3}}}(2)$, or vice-versa. In Fig.~\ref{fig:edgeN3}, we provide two different of set of parameters corresponding to these two possible ground-states, hence showing explicitly the inversion symmetry breaking.

\subsection{Entanglement spectrum in finite-size systems}
In the Letter, we have used the structure of the entanglement spectrum to identify the topological properties 
underlying the ground-state wave function.  However, our calculations were done for finite-size systems and 
it is not obvious to what extent the theoretical predictions, that are made using the properties of infinite-size 
matrix-product states \cite{Garcia-W-S-V-C-08,Pollmann-T-B-O-10,Tanimoto-T-14}, are valid.   
To illustrate how finite-size calculations work in getting getting the information on the topological properties, 
we calculate the entanglement spectrum (i.e., the Schmidt eigenvalues) of the $(\mathbf{3},\bar{\mathbf{3}})$ VBS state 
\cite{Affleck-K-L-T-88,Morimoto-U-M-F-14} discussed above.   
Following the standard procedure \cite{Shi-D-V-06}, we can obtain the entanglement spectrum for a finite-size ($L$) 
system:  
\begin{equation}
\frac{\sqrt{1+ 2(-1/8)^{\ell_{\text{L}}} } \sqrt{1 + 2(-1/8)^{\ell_{\text{R}}} }}{\sqrt{3} \sqrt{1 + 2(-1/8)^{L} }} \; , \;\;
\frac{\sqrt{1- (-1/8)^{\ell_{\text{L}}} } \sqrt{1 - (-1/8)^{\ell_{\text{R}}} }}{\sqrt{3} \sqrt{1 + 2(-1/8)^{L} }} \;\; (\times 2)  \; ,
\end{equation}
where $\ell_{\text{L}}$ and $\ell_{\text{R}}$ respectively are the sizes of the left and right subsystems 
($L=\ell_{\text{L}}+\ell_{\text{R}}$), and the edge states are fixed to $\lambda_{\text{\bf 3}}(1)$ (left) 
and $\lambda_{\bar{\text{\bf 3}}}(1)$ (right).    
For finite-size systems, the three-fold degeneracy, which is a clear signature of the topological property, 
is weakly broken (with an exponentially small splitting).   
Roughly, the fictitious SU(3) spins $\mathbf{3}$ and $\bar{\mathbf{3}}$ at the entanglement cut feel 
the (exponentially small) effects of the actual (emergent) edge spins on the boundaries thereby breaking 
the perfect degeneracy characteristic of free spins. 

\subsection{Quantum phase transition induced by $t_\perp$ in the $N=3$ model}

\begin{figure}[!ht]
\centering
\includegraphics[width=0.8\columnwidth,clip]{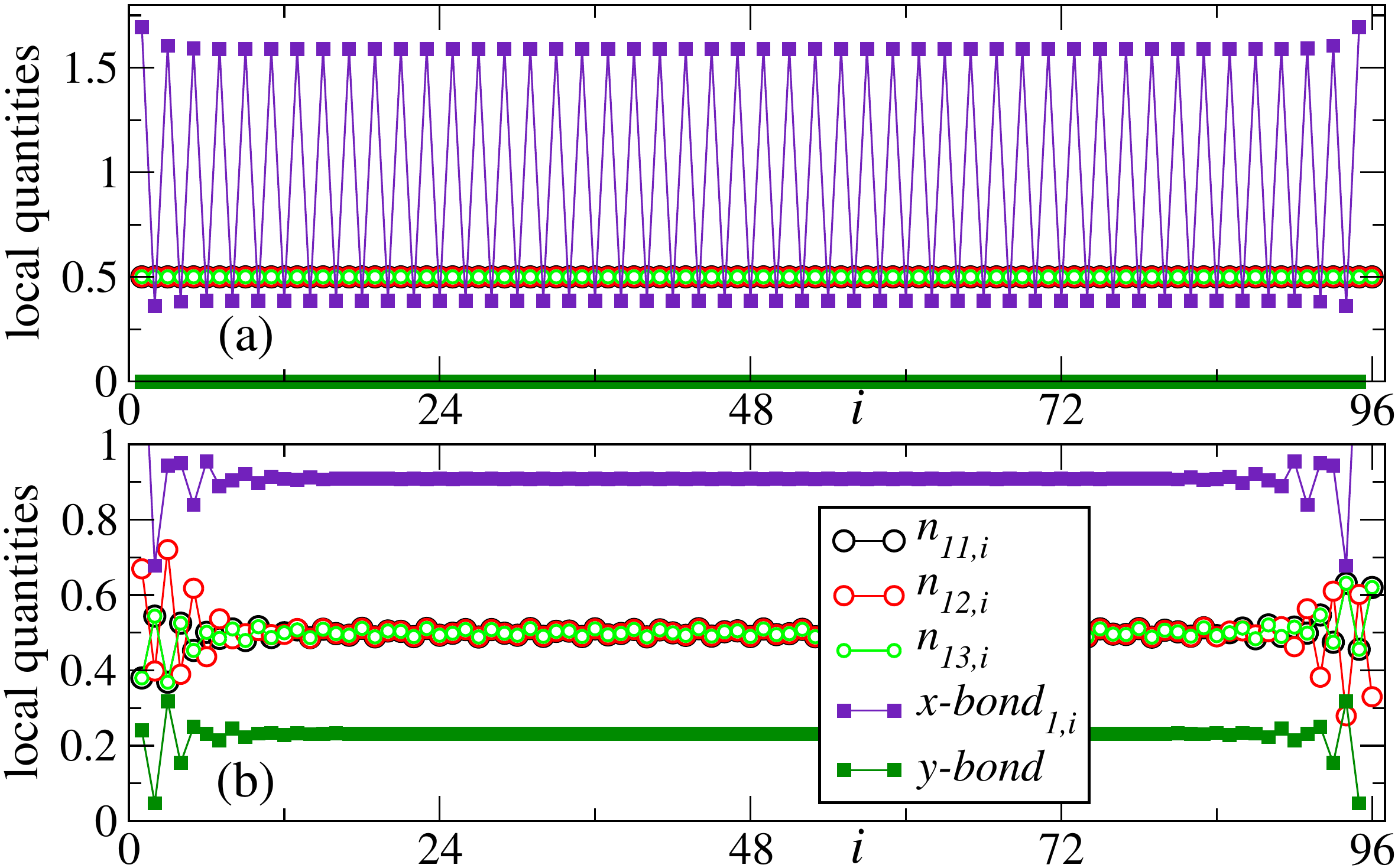}
\caption{Local quantities for the $N=3$ model obtained by DMRG simulations for $L=96$, $(U,V)=(100,10)$ and varying $t_\perp/t=0.4$ (a) and $1$ (b) respectively. }
\label{fig:QPT_N3}
\end{figure}

As discussed in the strong coupling section above, the effective model is highly non-trivial for $N=3$ and in particular, there is a crucial role played by $t_\perp$ which acts as an effective magnetic field for the $d$ orbitals. In Fig.~\ref{fig:QPT_N3c} and Fig.~\ref{fig:QPT_N3} , we plot the local quantities (see main text) for large interactions and various $t_\perp$, so that we can identify several phases. At small $t_\perp/t=0.4$, one clearly observes in-phase dimerization, corresponding to a uniform spin Peierls-like phase $0$-SP. When $t_\perp=t$, we recover the $N=3$ chiral SPT phase, which in this particular simulation corresponds to a left-edge having $\lambda_{\text{\bf 3}}(2)$ (hence a right-edge corresponding to $\lambda_{\bar{\text{\bf 3}}}(2)$). Finally, for large $t_\perp/t=10$, there is a quantum phase transition to a featureless fully gapped phase. Note that in this case, the rung energy is very close to $1$, as expected when orbital fluctuations are suppressed. 

%


\begin{thebibliography}{101}
\bibitem{wenbook}
B. Zeng, X. Chen, D. L. Zhou, and X. G. Wen, arXiv:1508.02595.

\bibitem{senthilreview}
T. Senthil, Annual Review of Condensed Matter Physics {\bf 6}, 299 (2015).

\bibitem{haldane}
F. D. M. Haldane, Phys. Lett. A {\bf 93}, 464 (1983); Phys. Rev. Lett. {\bf 50}, 1153 (1983).

\bibitem{Kennedy-90}
T.~Kennedy, J. Phys.: Condensed Matter, {\bf 2}, 5737 (1990).

\bibitem{hagiwara}
M. Hagiwara, K. Katsumata, I. Affleck, B. I. Halperin, and J. P. Renard,
Phys. Rev. Lett. {\bf 65}, 3181 (1990).

\bibitem{gu}
Z. C. Gu and X. G. Wen,  Phys. Rev. B {\bf 80}, 155131 (2009).

\bibitem{pollmann}
F. Pollmann, E. Berg, A. M. Turner, and M. Oshikawa,
Phys. Rev. B {\bf 85}, 075125 (2012).

\bibitem{chenx}
X. Chen, Z.-C. Gu, and X.-G. Wen,
Phys. Rev. B {\bf 83}, 035107 (2011); 
Phys. Rev. B {\bf 84}, 235128 (2011).

\bibitem{schuch}
N. Schuch, D. Perez-Garcia, and I. Cirac,
Phys. Rev. B {\bf 84}, 165139 (2011).

\bibitem{fidowski}
L. Fidkowski and A. Kitaev, Phys. Rev. B  {\bf 83}, 075103 (2011).

\bibitem{Pollmann2010}
F. Pollmann, A. M. Turner, E. Berg, and M. Oshikawa,
Phys. Rev. B {\bf 81}, 064439 (2010).

\bibitem{xu}
Z. Bi, A. Rasmussen, K. Slagle, and C. Xu,
Phys. Rev. B {\bf 91}, 134404 (2015).

\bibitem{Duivenvoorden-Q-13}
K. Duivenvoorden and T. Quella, Phys. Rev. B {\bf 87}, 125145 (2013).

\bibitem{comment}
As SU($N$) does not possess non-trivial projective representations, 
we need to consider the projective unitary group $\text{PSU($N$)} \cong  \text{SU($N$)}/\mathbb{Z}_N$ as the protecting symmetry. 

\bibitem{Else-B-D-13}
D.V.~Else, S.D.~Bartlett, and A.C.~Doherty, Phys. Rev. B {\bf 88}, 085114 (2013). 

\bibitem{Duivenvoorden-Q-ZnxZn-13}
K. Duivenvoorden and T. Quella, Phys. Rev. B {\bf 88}, 125115 (2013).

\bibitem{Capponi-L-T-15}
S. Capponi, P. Lecheminant, and K. Totsuka,
Ann. Phys. {\bf 367}, 50 (2016).

\bibitem{Furusaki2014}
T. Morimoto, H. Ueda, T. Momoi, and A. Furusaki,
Phys. Rev. B {\bf 90}, 235111 (2014).

\bibitem{greiter}
S. Rachel, D. Schuricht, B. Scharfenberger, R. Thomale, and M. Greiter,
J. Phys.: Conf. Ser. {\bf 200}, 022049 (2010).

\bibitem{AKLT}
I. Affleck, T. Kennedy, E.H. Lieb, H. Tasaki, Comm. Math. Phys. {\bf 115}, 477 (1988).

\bibitem{Quella2015}
A. Roy and T. Quella,  arXiv:1512.05229.

\bibitem{Cazalilla-H-U-09}
M. A. Cazalilla, A. F. Ho, and M. Ueda, New J. Phys. {\bf 11}, 103033 (2009).

\bibitem{Gorshkov-et-al-10}
A. V. Gorshkov, M. Hermele, V. Gurarie, C. Xu, P. S. Julienne, 
J. Ye, P. Zoller, E. Demler, M. D. Lukin, and A. M. Rey, 
Nat. Phys. {\bf 6}, 289 (2010).

\bibitem{Cazalilla-R-14}
M. A. Cazalilla and A. M. Rey, Rep. Prog. Phys. {\bf 77}, 124401
(2014).

\bibitem{Taie2012}
S. Taie, R. Yamazaki, S. Sugawa, and Y. Takahashi,
Nat. Phys. {\bf 8}, 825 (2012).

\bibitem{Pagano2014}
G. Pagano, M. Mancini, G. Cappellini, P. Lombardi, F. Schafer,
H. Hu, X.-J. Liu, J. Catani, C. Sias, M. Inguscio, and L. Fallani,
Nat. Phys. {\bf 10}, 198 (2014).

\bibitem{Zhang2014}
X. Zhang, M. Bishof, S. L. Bromley, C. V. Kraus, M. S.
Safronova, P. Zoller, A. M. Rey, and J. Ye, Science {\bf 345}, 1467
(2014).

\bibitem{Scazza2014}
F. Scazza, C. Hofrichter, M. H\"ofer, P. C. De Groot, I. Bloch,
and S. F\"olling, Nat. Phys. {\bf 10}, 779 (2014).

\bibitem{atala}
M. Atala, M. Aidelsburger, M. Lohse, J. T. Barreiro, B. Paredes,
and I. Bloch, Nat. Phys {\bf 10}, 588 (2014).

\bibitem{sebbystrabley}
J. Sebby-Strabley, M. Anderlini, P. S. Jessen, and J. V. Porto,
Phys. Rev. A {\bf 73}, 033605 (2006).

\bibitem{bloch:RMP2008}
I. Bloch, J. Dalibard, and W. Zwerger,
Rev. Mod. Phys. {\bf 80}, 885 (2008).

\bibitem{bloch:AAMO2006}
I. Bloch and M. Greiner, Adv. At. Mol. Opt. Phy. {\bf 52}, 1 (2006).

\bibitem{werner:PRL2005}
F. Werner, O. Parcollet, A. Georges, and S. R. Hassan,
Phys. Rev. Lett. {\bf 95}, 056401 (2005). 

\bibitem{note1}
Typically, increasing $r$ makes, e.g., $t_{\perp}/t$ and $V/U$ smaller.

\bibitem{chin:RMP2010}
  C. Chin, R. Grimm, P. Julienne, and E. Tiesinga,
  Rev. Mod. Phys. \textbf{82}, 1225 (2010).

\bibitem{enomoto:PRL2008}
  K. Enomoto, K. Kasa, M. Kitagawa, and Y. Takahashi,
  Phys. Rev. Lett. \textbf{101}, 203201 (2008).

\bibitem{taie:PRL2016}
  S. Taie, S. Watanabe, T. Ichinose,  and Y. Takahashi,
  Phys. Rev. Lett. \textbf{116}, 043202 (2016).

\bibitem{Kobayashi-O-O-Y-M-12}
K. Kobayashi, M. Okumura, Y. Ota, S. Yamada, and
M. Machida, Phys. Rev. Lett.  {\bf 109},  235302 (2012).

\bibitem{Kobayashi-O-O-Y-M-14}
K. Kobayashi, Y. Ota, M. Okumura, S. Yamada, and
M. Machida, Phys. Rev. A  {\bf 89}, 023625 (2014).

\bibitem{Bois-C-L-M-T-15}
V. Bois, S. Capponi, P. Lecheminant, M. Moliner, and K. Totsuka,
Phys. Rev. B {\bf 91},  075121 (2015).

\bibitem{supp} See the supplementary material for more information.

\bibitem{Nonne-M-C-L-T-13}
H. Nonne, M. Moliner, S. Capponi, P. Lecheminant, and K. Totsuka,
EPL {\bf 102},  37008 (2013).

\bibitem{Totsuka-15}
K. Tanimoto and K. Totsuka, arXiv:1508.07601.

\bibitem{Mila-16}
K. Wan, P. Nataf, and F. Mila, Phys. Rev. B {\bf 96}, 115159 (2017).

\bibitem{3-3bar-ladder} 
S.~Capponi, P.~Fromholz, P.~Lecheminant, and K.~Totsuka, in preparation. 

\bibitem{DMRG}
S. R. White, Phys. Rev. Lett. {\bf 69}, 2863 (1992);
U. Schollw{\"o}ck, Rev. Mod. Phys. {\bf 77}, 259 (2005).

\bibitem{Hasebe-T-13}
K.~Hasebe and K.~Totsuka, 
Phys. Rev. B {\bf 87}, 045115 (2013). 

\bibitem{Pollmann2015}
S. Moudgalya and F. Pollmann,
Phys. Rev. B  {\bf 91}, 155128 (2015).

\bibitem{Islam-et-al-entanglement-15}
R.~Islam, R.~Ma, P.M.~Preiss, M.~Eric Tai, A.~Lukin, M.~Rispoli, and M.~Greiner, 
Nature {\bf 528}, 77 (2015). 

\bibitem{zurn_PRL2013}
G. Zurn, T. Lompe, A. N. Wenz, S. Jochim, P. S. Julienne, and J. M. Hutson,
Phys. Rev. Lett. \textbf{110}, 135301 (2013).

\bibitem{boll_Science2016}
M. Boll, T. A. Hilker, G. Salomon, A. Omran, J. Nespolo,
L. Pollet, I. Bloch, and C. Gross,
Science \textbf{353}, 1257 (2016).

\bibitem{parsons_Science2016}
M. F. Parsons, A. Mazurenko, C. S. Chiu, G. Ji, D. Greif, and M. Greiner,
Science \textbf{353}, 1253 (2016).

\bibitem{Becker2017} 
J. Becker, T. K\"ohler, A. C. Tiegel, S. R. Manmana, S. Wessel, and A. Honecker,Phys. Rev. B {\bf 96}, 060403(R) (2017).

\bibitem{kitagawa:PRA2008}
M. Kitagawa, K. Enomoto, K. Kasa, Y. Takahashi, R. Ciury, P. Naidon, and P. S. Julienne,
Phys. Rev. A {\bf 77}, 012719 (2008).

\bibitem{hofrichter:PRX2016}
C. Hofrichter, L. Riegger, F. Scazza, M. H{\"o}fer, D. Rio Fernandes, I. Bloch, and 
S. F{\"o}lling, Phys. Rev. X {\bf 6}, 021030 (2016).

\bibitem{stellmer_PRA2011}
S. Stellmer, R. Grimm, and F. Schreck,
Phys. Rev. A \textbf{84}, 043611 (2011).




\end{thebibliography}

\begin{thebibliography}{10}%
\makeatletter
\providecommand \@ifxundefined [1]{%
 \@ifx{#1\undefined}
}%
\providecommand \@ifnum [1]{%
 \ifnum #1\expandafter \@firstoftwo
 \else \expandafter \@secondoftwo
 \fi
}%
\providecommand \@ifx [1]{%
 \ifx #1\expandafter \@firstoftwo
 \else \expandafter \@secondoftwo
 \fi
}%
\providecommand \natexlab [1]{#1}%
\providecommand \enquote  [1]{``#1''}%
\providecommand \bibnamefont  [1]{#1}%
\providecommand \bibfnamefont [1]{#1}%
\providecommand \citenamefont [1]{#1}%
\providecommand \href@noop [0]{\@secondoftwo}%
\providecommand \href [0]{\begingroup \@sanitize@url \@href}%
\providecommand \@href[1]{\@@startlink{#1}\@@href}%
\providecommand \@@href[1]{\endgroup#1\@@endlink}%
\providecommand \@sanitize@url [0]{\catcode `\\12\catcode `\$12\catcode
  `\&12\catcode `\#12\catcode `\^12\catcode `\_12\catcode `\%12\relax}%
\providecommand \@@startlink[1]{}%
\providecommand \@@endlink[0]{}%
\providecommand \url  [0]{\begingroup\@sanitize@url \@url }%
\providecommand \@url [1]{\endgroup\@href {#1}{\urlprefix }}%
\providecommand \urlprefix  [0]{URL }%
\providecommand \Eprint [0]{\href }%
\providecommand \doibase [0]{http://dx.doi.org/}%
\providecommand \selectlanguage [0]{\@gobble}%
\providecommand \bibinfo  [0]{\@secondoftwo}%
\providecommand \bibfield  [0]{\@secondoftwo}%
\providecommand \translation [1]{[#1]}%
\providecommand \BibitemOpen [0]{}%
\providecommand \bibitemStop [0]{}%
\providecommand \bibitemNoStop [0]{.\EOS\space}%
\providecommand \EOS [0]{\spacefactor3000\relax}%
\providecommand \BibitemShut  [1]{\csname bibitem#1\endcsname}%
\let\auto@bib@innerbib\@empty
\bibitem [{\citenamefont {Itzykson}\ and\ \citenamefont
  {Nauenberg}(1966)}]{Itzykson-N-66}%
  \BibitemOpen
  \bibfield  {author} {\bibinfo {author} {\bibfnamefont {C.}~\bibnamefont
  {Itzykson}}\ and\ \bibinfo {author} {\bibfnamefont {M.}~\bibnamefont
  {Nauenberg}},\ }\href {http://link.aps.org/doi/10.1103/RevModPhys.38.95}
  {\bibfield  {journal} {\bibinfo  {journal} {Rev. Mod. Phys.}\ }\textbf
  {\bibinfo {volume} {38}},\ \bibinfo {pages} {95} (\bibinfo {year}
  {1966})}\BibitemShut {NoStop}%
\bibitem [{\citenamefont {Georgi}(1999)}]{Georgi-book-99}%
  \BibitemOpen
  \bibfield  {author} {\bibinfo {author} {\bibfnamefont {H.}~\bibnamefont
  {Georgi}},\ }\href@noop {} {\emph {\bibinfo {title} {Lie Algebras in Particle
  Physics}}}\ (\bibinfo  {publisher} {Perseus Books},\ \bibinfo {year}
  {1999})\BibitemShut {NoStop}%
\bibitem [{\citenamefont {Bois}\ \emph {et~al.}(2015)\citenamefont {Bois},
  \citenamefont {Capponi}, \citenamefont {Lecheminant}, \citenamefont
  {Moliner},\ and\ \citenamefont {Totsuka}}]{Bois-C-L-M-T-15supp}%
  \BibitemOpen
  \bibfield  {author} {\bibinfo {author} {\bibfnamefont {V.}~\bibnamefont
  {Bois}}, \bibinfo {author} {\bibfnamefont {S.}~\bibnamefont {Capponi}},
  \bibinfo {author} {\bibfnamefont {P.}~\bibnamefont {Lecheminant}}, \bibinfo
  {author} {\bibfnamefont {M.}~\bibnamefont {Moliner}}, \ and\ \bibinfo
  {author} {\bibfnamefont {K.}~\bibnamefont {Totsuka}},\ }\href {\doibase
  10.1103/PhysRevB.91.075121} {\bibfield  {journal} {\bibinfo  {journal} {Phys.
  Rev. B}\ }\textbf {\bibinfo {volume} {91}},\ \bibinfo {pages} {075121}
  (\bibinfo {year} {2015})}\BibitemShut {NoStop}%
\bibitem [{\citenamefont {Affleck}\ \emph {et~al.}(1988)\citenamefont
  {Affleck}, \citenamefont {Kennedy}, \citenamefont {Lieb},\ and\ \citenamefont
  {Tasaki}}]{Affleck-K-L-T-88}%
  \BibitemOpen
  \bibfield  {author} {\bibinfo {author} {\bibfnamefont {I.}~\bibnamefont
  {Affleck}}, \bibinfo {author} {\bibfnamefont {T.}~\bibnamefont {Kennedy}},
  \bibinfo {author} {\bibfnamefont {E.~H.}\ \bibnamefont {Lieb}}, \ and\
  \bibinfo {author} {\bibfnamefont {H.}~\bibnamefont {Tasaki}},\ }\href
  {http://dx.doi.org/10.1007/BF01218021} {\bibfield  {journal} {\bibinfo
  {journal} {Comm. Math. Phys.}\ }\textbf {\bibinfo {volume} {115}},\ \bibinfo
  {pages} {477} (\bibinfo {year} {1988})},\ \bibinfo {note}
  {10.1007/BF01218021}\BibitemShut {NoStop}%
\bibitem [{\citenamefont {Morimoto}\ \emph {et~al.}(2014)\citenamefont
  {Morimoto}, \citenamefont {Ueda}, \citenamefont {Momoi},\ and\ \citenamefont
  {Furusaki}}]{Morimoto-U-M-F-14}%
  \BibitemOpen
  \bibfield  {author} {\bibinfo {author} {\bibfnamefont {T.}~\bibnamefont
  {Morimoto}}, \bibinfo {author} {\bibfnamefont {H.}~\bibnamefont {Ueda}},
  \bibinfo {author} {\bibfnamefont {T.}~\bibnamefont {Momoi}}, \ and\ \bibinfo
  {author} {\bibfnamefont {A.}~\bibnamefont {Furusaki}},\ }\href {\doibase
  10.1103/PhysRevB.90.235111} {\bibfield  {journal} {\bibinfo  {journal} {Phys.
  Rev. B}\ }\textbf {\bibinfo {volume} {90}},\ \bibinfo {pages} {235111}
  (\bibinfo {year} {2014})}\BibitemShut {NoStop}%
\bibitem [{\citenamefont {Duivenvoorden}\ and\ \citenamefont
  {Quella}(2013)}]{Duivenvoorden-Q-13supp}%
  \BibitemOpen
  \bibfield  {author} {\bibinfo {author} {\bibfnamefont {K.}~\bibnamefont
  {Duivenvoorden}}\ and\ \bibinfo {author} {\bibfnamefont {T.}~\bibnamefont
  {Quella}},\ }\href {http://link.aps.org/doi/10.1103/PhysRevB.87.125145}
  {\bibfield  {journal} {\bibinfo  {journal} {Phys. Rev. B}\ }\textbf {\bibinfo
  {volume} {87}},\ \bibinfo {pages} {125145} (\bibinfo {year}
  {2013})}\BibitemShut {NoStop}%
\bibitem [{\citenamefont {P{\'{e}}rez-Garc{\'{i}}a}\ \emph
  {et~al.}(2008)\citenamefont {P{\'{e}}rez-Garc{\'{i}}a}, \citenamefont {Wolf},
  \citenamefont {Sanz}, \citenamefont {Verstraete},\ and\ \citenamefont
  {Cirac}}]{Garcia-W-S-V-C-08}%
  \BibitemOpen
  \bibfield  {author} {\bibinfo {author} {\bibfnamefont {D.}~\bibnamefont
  {P{\'{e}}rez-Garc{\'{i}}a}}, \bibinfo {author} {\bibfnamefont {M.~M.}\
  \bibnamefont {Wolf}}, \bibinfo {author} {\bibfnamefont {M.}~\bibnamefont
  {Sanz}}, \bibinfo {author} {\bibfnamefont {F.}~\bibnamefont {Verstraete}}, \
  and\ \bibinfo {author} {\bibfnamefont {J.~I.}\ \bibnamefont {Cirac}},\ }\href
  {http://link.aps.org/doi/10.1103/PhysRevLett.100.167202} {\bibfield
  {journal} {\bibinfo  {journal} {Phys. Rev. Lett.}\ }\textbf {\bibinfo
  {volume} {100}},\ \bibinfo {pages} {167202} (\bibinfo {year}
  {2008})}\BibitemShut {NoStop}%
\bibitem [{\citenamefont {Pollmann}\ \emph {et~al.}(2010)\citenamefont
  {Pollmann}, \citenamefont {Turner}, \citenamefont {Berg},\ and\ \citenamefont
  {Oshikawa}}]{Pollmann-T-B-O-10}%
  \BibitemOpen
  \bibfield  {author} {\bibinfo {author} {\bibfnamefont {F.}~\bibnamefont
  {Pollmann}}, \bibinfo {author} {\bibfnamefont {A.~M.}\ \bibnamefont
  {Turner}}, \bibinfo {author} {\bibfnamefont {E.}~\bibnamefont {Berg}}, \ and\
  \bibinfo {author} {\bibfnamefont {M.}~\bibnamefont {Oshikawa}},\ }\href
  {http://link.aps.org/doi/10.1103/PhysRevB.81.064439} {\bibfield  {journal}
  {\bibinfo  {journal} {Phys. Rev. B}\ }\textbf {\bibinfo {volume} {81}},\
  \bibinfo {pages} {064439} (\bibinfo {year} {2010})}\BibitemShut {NoStop}%
\bibitem [{\citenamefont {Tanimoto}\ and\ \citenamefont
  {Totsuka}()}]{Tanimoto-T-14}%
  \BibitemOpen
  \bibfield  {author} {\bibinfo {author} {\bibfnamefont {K.}~\bibnamefont
  {Tanimoto}}\ and\ \bibinfo {author} {\bibfnamefont {K.}~\bibnamefont
  {Totsuka}},\ }\href@noop {} {}\bibinfo {note} {ArXiv:1508.07601}\BibitemShut
  {NoStop}%
\bibitem [{\citenamefont {Shi}\ \emph {et~al.}(2006)\citenamefont {Shi},
  \citenamefont {Duan},\ and\ \citenamefont {Vidal}}]{Shi-D-V-06}%
  \BibitemOpen
  \bibfield  {author} {\bibinfo {author} {\bibfnamefont {Y.-Y.}\ \bibnamefont
  {Shi}}, \bibinfo {author} {\bibfnamefont {L.-M.}\ \bibnamefont {Duan}}, \
  and\ \bibinfo {author} {\bibfnamefont {G.}~\bibnamefont {Vidal}},\ }\href
  {http://link.aps.org/doi/10.1103/PhysRevA.74.022320} {\bibfield  {journal}
  {\bibinfo  {journal} {Phys. Rev. A}\ }\textbf {\bibinfo {volume} {74}},\
  \bibinfo {pages} {022320} (\bibinfo {year} {2006})}\BibitemShut {NoStop}%
\end{thebibliography}
\end{document}